\documentclass[tightenlines,showpacs,prd,floatfix,preprintnumbers,amsmath,amssymb,nofootinbib,superscriptaddress,twocolumn]{revtex4-1}
\DeclareMathAlphabet{\mathpzc}{OT1}{pzc}{m}{it}
\usepackage{graphicx}
\usepackage{amsfonts}
\usepackage{caption}
\usepackage{verbatim}
\usepackage{hyperref}
\usepackage[normalem]{ulem}
\usepackage{color}
\usepackage{mathrsfs}
\usepackage{bm}
\usepackage{comment}
\usepackage{ulem}

\def\up#1{\raise1mm\hbox{$\!\!^{#1}$}} 
\def\upp#1{\raise2mm\hbox{$\!\!\!\!^{#1}$}}

\newcommand{\beq}{\begin{equation}}
\newcommand{\be}{\begin{equation}}
\newcommand{\eeq}{\end{equation}}
\newcommand{\ee}{\end{equation}}
\newcommand{\bea}{\begin{eqnarray}}
\newcommand{\eea}{\end{eqnarray}}
\newcommand{\bal}{\begin{align}}
\newcommand{\eal}{\end{align}}

\newcommand{\beaa}{\begin{eqnarray*}} 
\newcommand{\eeaa}{\end{eqnarray*}}

\newcommand{\Lie}{\mbox{\pounds}}
\newcommand{\bsube}{\begin{subequations}}
\newcommand{\esube}{\end{subequations}}

\begin{document}

\title{Linear-in-mass-ratio contribution to spin precession and tidal invariants in Schwarzschild spacetime at very high post-Newtonian order}

\author{Abhay G. Shah} 
\email{a.g.shah@soton.ac.uk}
\affiliation{Mathematical Sciences, University of Southampton, Southampton SO17 1BJ, United Kingdom}
\author{Adam Pound} 
\email{a.pound@soton.ac.uk}
\affiliation{Mathematical Sciences, University of Southampton, Southampton SO17 1BJ, United Kingdom}
\date{\today}


\begin{abstract}
Using black hole perturbation theory and arbitrary-precision computer algebra, we obtain the post-Newtonian (pN) expansions of the linear-in-mass-ratio corrections to the spin-precession angle and tidal invariants for a particle in circular orbit around a Schwarzschild black hole.
%
%
%
We extract coefficients up to 20pN order from numerical results that are calculated with an accuracy greater than 1 part in $10^{500}$. 
These results can be used to calibrate parameters in effective-one-body models of compact binaries, specifically the spin-orbit part of the effective Hamiltonian and the dynamically significant tidal part of the main radial potential of the effective metric. 
Our calculations are performed in a radiation gauge, which is known to be singular away from the particle. To overcome this irregularity, we define suitable Detweiler-Whiting singular and regular fields in this gauge, and we compute the invariants using mode-sum regularization in combination with averaging from two sides of the particle. The detailed justification of this regularization procedure will be presented in a forthcoming companion paper. 
\end{abstract}

\maketitle

\section{Introduction}
\label{sec1}
Compact binaries are expected to be among the most significant sources for gravitational wave detectors, and over the last decade, modeling of these binaries has become a mature field. For two widely separated compact objects, post-Newtonian (pN) models of very high order have been developed~\cite{Blanchet:14,Damour-etal:14}. For compact objects of very dissimilar masses, calculations using gravitational self-force theory~\cite{Poisson-Pound-Vega:11,Barack:09} can now be performed with high accuracy at linear order in the mass ratio~\cite{Barack-Sago:10,Akcay-Warburton-Barack:13,Osburn-etal:14,Dolanetal-tidal,sf5}, and calculations at second order in the mass ratio are underway~\cite{Detweiler:12,Pound:12a,Gralla:12,Pound:14c,Pound-Miller:14}. For objects of similar mass and small separations, fully nonlinear numerical relativity can be used, and it is continually pushing to cover more of the binary parameter space~\cite{Szilagyi-etal:15}.  Furthermore, effective-one-body theory (EOB)~\cite{Buonanno:1998gg,Damour:2000we} has the potential to model all of these situations, given appropriate calibration data from each of the other models.

In the last few years, work has increasingly been done at the interface between these various models~\cite{LeTiec:14}. Of particular interest to us is that self-force data has been used to set benchmarks for numerical relativity, determine high-order pN parameters, and calibrate EOB. One thing that has aided these efforts is the advent of extreme-accuracy numerical self-force results, with accuracies of several hundred or thousands of digits~\cite{SKerr, sf5}. With these results, one can extract very high-order (linear-in-mass-ratio) post-Newtonian (pN) coefficients, which would be prohibitively difficult to calculate in an analytical pN framework. Perhaps surprisingly, with high-accuracy numerical self-force data one can extract not only numerical values of the pN coefficients, but their analytical form in terms of rational numbers (and their product with $\pi$), and for some terms, even analytical forms involving linear combinations of transcendentals. The details of this procedure are given in Ref.~\cite{Nathanetal}.

To use data from one model in another, one must calculate gauge- and coordinate-invariant quantities that can be translated between formalisms. These invariants can be divided into those related to dissipation and those related to conservative effects. In the dissipative sector, for example, Ref.~\cite{SKerr} calculated the ingoing-at-horizon and outgoing-at-infinity gravitational-wave fluxes to 21pN order for a particle in circular, equatorial orbit around a Kerr black hole. However, most of the work done has been in the conservative sector, where a richer variety of invariants appear. The paragon of conservative invariants is Detweiler's redshift invariant~\cite{detweiler08}, which is related to the binding energy and angular momentum of a quasi-circular binary system. Reference~\cite{sf5} determined the pN expansion of this invariant up to 10.5pN order for a particle in circular orbit around a Schwarzschild black hole, and Ref.~\cite{Nathanetal} has pushed that expansion even further, to 21.5pN order.

Recently, other conservative invariants on circular orbits in Schwarzschild spacetime have been discovered and calculated: the linear-in-mass-ratio correction to the spin precession (due to the ``self-torque'' effect) was numerically calculated in Ref.~\cite{Dolanetal-spin},  where the pN expansion was found up to 3pN order (also see \cite{BMFB}); and conservative tidal invariants, the eigenvalues of certain electric and magnetic tidal tensors, were calculated in Ref.~\cite{Dolanetal-tidal} to more than machine precision. 

In parallel with this numerical work, there have been analytical calculations of invariants, made possible by performing a pN expansion of the equations of self-force theory. Reassuringly, in all cases the results of the analytical calculations have precisely agreed with the expansions extracted from numerical data. Reference~\cite{Fujita-Kerr} analytically calculated the pN series of ingoing and outgoing fluxes for a particle in circular, equatorial orbit around a Kerr black hole and confirmed the results in Ref.~\cite{SKerr} to 11pN order. Bini $\&$ Damour  have analytically calculated the pN expansion of the redshift invariant up to 9.5pN order~\cite{BD1, BD2, BD3, BD4},  the spin precession  to 10.5pN order~\cite{BD-spin,BD-spin-new}, and tidal invariants to 7.5pN order~\cite{BD-tidal}. 

Self-force calculations, whether numerical or analytical, of these invariants have proven to be very effective in calibrating EOB. Bini $\&$ Damour have used the spin-precession invariant to calculate the corresponding terms in the spin-orbit part of the effective Hamiltonian through an 
effective gyro-gravitomagnetic ratio. 
They have likewise transcribed the tidal invariants into the tidal 
interaction energy in the EOB formalism. A semi-analytical simulation of tidally interacting binary neutron stars using these calibrations was performed in Ref.~\cite{Bernuzzietal}, and for compact stars the model was confirmed by fully numerical simulation in Ref.~\cite{Hotokezakaetal}. 

Despite these successes, at this point in time, computations of the spin-precession and tidal invariants have been performed at much lower accuracy than the extreme-accuracy calculations in Refs.~\cite{SKerr, sf5}. In this paper, we pursue the program initiated in Refs.~\cite{SKerr, sf5} by calculating the spin-precession and tidal invariants with an accuracy greater than 1 part in $10^{500}$ for circular orbits in Schwarzschild with orbital radii in the range $10^{18}$--$9\times10^{33}M$. This allows us to extract the pN expansion of the invariants  to about 20 pN order. Such a procedure of extracting pN coefficients from high precision numerical data will be very important when calculating conservative invariants for orbits in Kerr spacetime, where analytical checks or computations are likely to be extremely difficult. 

Besides our high accuracy, our calculations differ from others in our choice of gauge. Dolan \emph{et al}~\cite{Dolanetal-spin,Dolanetal-tidal} performed their calculations in the Lorenz gauge and Regge-Wheeler-Zerilli gauge. Bini $\&$ Damour~\cite{BD-spin, BD-tidal} performed theirs in the Regge-Wheeler-Zerilli gauge using analytical solutions of the Regge-Wheeler equation given by Ref.~\cite{MSTrw}. In this paper we use a radiation gauge, following the methods of Refs.~\cite{sf2,sf3,sf4,sf5}. Perturbations in radiation gauges can be reconstructed from solutions to the Teukolsky equation, making this gauge, unlike Lorenz and Regge-Wheeler-Zerilli, ideal for calculations in Kerr spacetime; hence, we view the calculations in this paper as a step toward computing analogous invariants in Kerr. 

Although it has this distinct advantage, the radiation gauge has historically had two drawbacks. First, the reconstructed metric does not contain the full solution of the linearized Einstein equation, but must be completed by adding a (non-radiation-gauge) perturbation that encodes the particle's mass and angular momentum. (We shall refer to the gauge of the completed solution as a completed radiation gauge; elsewhere, it has been called a modified radiation gauge.) Second, most quantities of interest in self-force theory are constructed from a certain smooth vacuum perturbation called the Detweiler-Whiting (DW) regular field~\cite{Detweiler:2002mi}, which in practice is obtained from the retarded metric perturbation by subtracting a certain divergent field called the DW singular field. Until recently, there has been no well-justified way of performing this procedure in a radiation gauge, which has long been known to have pathological singularities away from the particle~\cite{Barack-Ori:01}. 

However, both of these issues have, to a large extent, been dealt with. Regarding the completion problem, it has been incrementally better understood over the last few years~\cite{lpthesis,kfw,sf4,Sano:2014maa,Pound-Merlin-Barack:14}. The present paper provides some description of the requirements on the completion term in the metric perturbation in Schwarzschild, and a future paper will contain a fuller description of the problem and its partial resolution in Kerr~\cite{Merlin-etal:15}. Regarding the problem of regularization, recent work of Pound, Merlin, and Barack (hereafter PMB)~\cite{Pound-Merlin-Barack:14} has provided a rigorous means of obtaining the correct gravitational self-force from radiation-gauge perturbations, and this method has been concretely implemented in Ref.~\cite{MS}. Here we use a regularization scheme, based on the results of PMB, that combines standard mode-sum regularization~\cite{Barack-Ori:00} with a certain averaging procedure. The rigorous justification of this scheme will be presented in a companion paper, but here we go some ways toward making it sensible by providing a suitable definition of the DW regular field in our completed radiation gauge.

The article is organized as follows. In Sec.~\ref{sec2} we describe the setup of our calculations, the invariants we calculate, and our procedure for reconstructing and completing the retarded metric perturbation. In Sec.~\ref{sec3} we describe our regularization procedure. In Sec.~\ref{sec4}, we discuss the extraction of pN coefficients. Most of these coefficients we determine as numerical values, but we also determine many coefficients in analytical form. Because of the length of the expansions, we present the analytical coefficients in an appendix~\ref{appB}, and we present the numerical coefficients in the accompanying supplementary files.  We work in units with $G=c=1$.

Before proceeding to the body of the paper, we note that in parallel with our efforts, a group comprising Chris Kavanagh, Adrian Ottewill and Barry Wardell~\cite{kow} have also performed very high-order pN expansions of the same invariants, using analytical methods in the Regge-Wheeler-Zerilli gauge. At all orders that we have compared, these two independent calculations have yielded identical results.

\section{Setup}
\label{sec2}

\subsection{Circular orbits and invariant quantities}
We consider a particle of mass $\mu$ orbiting a Schwarzschild black hole of mass $M$. At zeroth order in $\mu/M$, the  trajectory is a circular geodesic $\gamma_0$ of the Schwarzschild metric $g_{\mu\nu}$. In Schwarzschild coordinates, its 4-velocity is given by
\begin{align} \allowdisplaybreaks
u^\alpha &= u^t (t^\alpha + \Omega \phi^\alpha),\,\textrm{with} \nonumber \\
u^t &= \frac{1}{ \sqrt{ 1-\frac{3M}{r_0}} },\, \textrm{and} \nonumber  \\
\Omega &= \sqrt{\frac{M} {r_0^3}},
\end{align}
where $r_0$ is the orbital radius, $\Omega$ is the orbital frequency, and $\phi^\alpha$ and $t^\alpha$ are the rotational and timelike Killing vectors of the Schwarzschild spacetime, respectively. The orbital energy and angular momentum per unit mass are given by 
\bea
\hat{E} &=&  \frac{ 1 - \frac{2M}{r_0} } { \sqrt{1-\frac{3M}{r_0}}  } , \nonumber \\
\hat{L}_z &=& \sqrt{ \frac{M r_0}{1 - \frac{3M}{r_0}} }.
\eea

At linear order in $\mu/M$, the particle sources a perturbation $h_{\mu\nu}$ that exerts a self-force on the particle. The perturbed trajectory $\gamma$ is no longer a geodesic of the background metric, but it is a geodesic of an \emph{effective metric} $\tilde g_{\mu\nu}=g_{\mu\nu}+h^R_{\mu\nu}$, where $h^R_{\mu\nu}$ is the DW regular field. If we consider only the conservative dynamics, the perturbed trajectory can be chosen, like the zeroth-order one, to be a circular orbit. 

The perturbed spacetime loses the background's stationarity and rotational symmetry, but it inherits the orbit's helical symmetry. In convenient gauges, the helical Killing vector is given by 
\beq\label{k}
k^\mu = t^\alpha + \Omega \phi^\alpha, 
\eeq
and at all points in the spacetime we have $\Lie_k h_{\mu\nu}=0=\Lie_k h^R_{\mu\nu}$. We can use this Killing vector in combination with $h^R_{\mu\nu}$ or derivatives of $h^R_{\mu\nu}$ to construct meaningful scalar quantities on the orbit. Let $\tilde u^\mu$ be the four-velocity of the perturbed trajectory, normalized as $\tilde g_{\mu\nu}\tilde u^\mu\tilde u^\nu=-1$. The simplest of the invariants is Detweiler's redshift invariant,
\beq\label{ut}
\tilde u^t = 1/\sqrt{-\tilde g_{\mu\nu}k^\mu k^\nu}|_\gamma,
\eeq
which is the rate of change of the time of an inertial frame at infinity relative to proper time in $\tilde g_{\mu\nu}$ along the orbit. Next is the spin precession invariant,
\beq\label{psi}
\tilde\psi = 1 - \tilde \omega/\tilde u^\phi, 
\eeq
which for a spinning particle is the angle of spin precession per unit radian of the circular orbit. Here the precession frequency is $\omega^2=-\frac{1}{2}(\tilde u^t)^2\tilde K_\mu{}^\nu\tilde K_\nu{}^\mu$, where $\tilde K_\mu{}^\nu=\tilde\nabla_\mu k^\nu|_\gamma$, and $\tilde\nabla_\mu$ is the covariant derivative compatible with $\tilde g_{\mu\nu}$. The last set of quantities we shall consider are the eigenvalues of the electric and magnetic quadrupole tidal tensors associated with $\tilde g_{\mu\nu}$. These are conveniently written in terms of a Lie-transported spatial triad $(\tilde e^\mu_1,\tilde e^\mu_2,\tilde e^\mu_3)$ on $\gamma$. The eigenvalues of the electric tidal tensor are
\begin{align}\label{lambdaE}
\tilde\lambda^\textrm{E}_i &= \tilde R_{\mu\alpha\nu\beta}\tilde e^\mu_i \tilde u^\alpha \tilde e^\nu_i \tilde u^\beta,
\end{align}
(with no summation over $i$); the magnetic tidal tensor has two equal but opposite eigenvalues, with magnitude
\beq\label{lambdaB}
\tilde\lambda^\textrm{B} = \tilde R_{\mu\alpha\nu\beta}\tilde u^\mu\tilde e^\alpha_2 \tilde e^\nu_2 \tilde e^\beta_3.
\eeq
Here $\tilde R_{\mu\alpha\nu\beta}$ is the Riemann tensor of $\tilde g_{\mu\nu}$.

From these scalars, we can obtain gauge-invariant perturbative quantities by expanding $\tilde g_{\mu\nu}$ around $g_{\mu\nu}$ and $\gamma$ around $\gamma_0$. 
Of interest to us in this paper are the linear perturbations in $\tilde\psi$, $\tilde\lambda^\textrm{E}_i$, and $\tilde\lambda^\textrm{B}$, which  were derived in Refs.~\cite{Dolanetal-spin} and \cite{Dolanetal-tidal}. The explicit expressions in Schwarzschild coordinates are
\begin{widetext}
\begin{align} \allowdisplaybreaks
\Delta\psi &= \frac{h^R_{tr,\phi}-h^R_{t\phi,r}+\Omega\left( h^R_{r\phi,\phi} - h_{\phi\phi,r} + f_0 r_0 h^R_{rr} \right)}{2r_0u^\phi} + \frac{u^\phi \left[ \Omega \left( M r_0^2 h_{tt} + r_0 f_0^2h^R_{\phi\phi} \right) + 2Mf_0h^R_{t\phi}\right]}{2 M f_0}, \label{Deltapsi} \\
\Delta\lambda_1^\textrm{E} &= \frac{\Omega^2 f_0 \left(2r_0-3M \right)}{r_0-3M}h_{rr} - \frac{ \Omega^2 \left(2r_0^2-6Mr_0+3M^2\right)}{f_0 \left( r_0-3M\right)^2}h^R_{tt} - \frac{6M\Omega f_0}{r_0\left(r_0-3M\right)^2}h^R_{t\phi} - \frac{\Omega^2\left(r_0^2-3Mr_0+3M^2\right)}{r_0^2\left(r_0-3M\right)^2}h^R_{\phi\phi} \nonumber \\ &- \frac{r_0-2M}{2(r_0-3M)}h^R_{kk,rr} - \frac{\Omega}{r_0} h^R_{rk,\phi} - \frac{1}{r_0} h^R_{tk,r}, \label{DeltalambdaE1}\\
\Delta\lambda_2^\textrm{E} &= \frac{M}{r_0(r_0-3M)^2}h^R_{kk} - \frac{1}{2r_0(r_0-3M)}h^R_{kk,\theta\theta} - \frac{\Omega^2}{r_0(r_0-3M)}h^R_{\theta\theta}, \\
\Delta\lambda_3^\textrm{E} &= \frac{\Omega^2}{f_0}h^R_{tt} - \Omega^2 f_0 \, h^R_{rr} - \frac{\Omega^2}{r_0^2} h^R_{\phi\phi} + \frac{\Omega}{r_0} h^R_{k\phi,r} - \frac{\Omega}{r_0} h^R_{rk,\phi} - \frac{1}{2r_0^2f_0}h^R_{kk,\phi\phi}, \displaybreak\\
\Delta\lambda^\textrm{B} &= \frac{3\Omega^3f_0^{1/2}}{(r_0-3M)} h^R_{\theta\theta} + \frac{\Omega^3f_0^{1/2}(r_0-9M)}{2(r_0-3M)^2} h^R_{\phi\phi} - \frac{\Omega^2(r_0-M)}{f_0^{1/2}(r_0-3M)^2} h^R_{t\phi} - \frac{\Omega M (5r_0-9M)}{2f_0^{1/2}r_0(r_0-3M)^2} h_{tt} - \frac{\Omega f_0^{3/2}}{r_0} h^R_{rr} \nonumber \\
&+ \frac{f_0^{1/2}}{2r_0^2} \left[ \Omega\left[  h^R_{\theta\theta,r} - 2h^R_{r\theta,\theta}  - h^R_{r\phi,\phi} \right] - h^R_{tr,\phi}\right] + \frac{1}{2f_0^{1/2}r_0^3} \left[ (r_0-4M) h^R_{t\phi,r} + \Omega(r_0-3M)h^R_{\phi\phi,r} \right] - \frac{\Omega}{2f_0^{1/2}}h^R_{tt,r} \nonumber \\
&+ \frac{\Omega}{2f_0^{1/2}r_0^2(r_0-3M)} \left[ f_0\, h^R_{\phi\phi,\theta\theta}+r_0^2\, h^R_{tt,\theta\theta} \right] - \frac{1}{2f_0^{1/2}r_0^3} h^R_{k\theta,\theta\phi} + \frac{(r_0-M)}{2f_0^{1/2}r_0^3(r_0-3M)} h^R_{t\phi,\theta\theta},\label{DeltalambdaB}
\end{align}
\end{widetext}
where $f_0=1-2M/r_0$, $h_{kk}=h_{ab}k^a k^b$, all functions are evaluated on the zeroth-order trajectory, and commas denote partial derivatives with respect to coordinates. The tidal invariants are related by $\Delta\lambda_1^\textrm{E}+\Delta\lambda_2^\textrm{E}+\Delta\lambda_3^\textrm{E}=0$.

The quantities $\Delta \psi$, $\lambda_i^{\rm E}$, and $\lambda^{\rm B}$, as given above, are defined by performing an expansion at fixed $\Omega$. They are invariant under any smooth gauge transformation that leaves the helical Killing vector invariant; that is, for a smooth gauge generator $\xi^\mu$, the quantities are invariant if $\Lie_\xi k^\mu=0$ (and hence $\Lie_k \xi^\mu=0$). In practice, this means that to obtain the same result in any two gauges, we must ensure that the perturbation $h_{\mu\nu}$ in each of these gauges possesses three properties: (i) helical symmetry, $\Lie_k h_{\mu\nu}=0$, with $k^\mu$ taking the form~\eqref{k}, (ii) asymptotic flatness, (iii) regularity at $\theta=0$ and $\pi$, and (iv) continuity of certain metric components. We explain these requirements in the next section.

\subsection{Computation of the retarded field}
We calculate the retarded field $h_{\mu\nu}$ in a completed radiation gauge. After decomposing the perturbation into tensor harmonic $\ell$ modes, we calculate the $\ell \ge 2$ modes in an outgoing radiation gauge satisfying $h_{ab}l^a=0$, where $l^a$ is the outgoing principal null vector. We calculate the $\ell=0,1$ modes, which physically correspond to the spacetime's change in mass and angular momentum, in the asymptotically flat Zerilli gauge. We briefly review the reconstruction formalism here; more detailed descriptions are available in Refs.~\cite{sf2, sf3, sf4, sf5, MS}. For the completion, we make several points about continuity which have not previously been discussed in the literature.

\subsubsection{Reconstruction}
The $\ell\geq2$ part of the metric perturbation is extracted from a Hertz potential, $\Psi$, which in turn is calculated from the perturbed spin-2 Weyl scalar, $\psi_0$, by solving the separable Bardeen-Press-Teukolsky equation. The spin-2 retarded Weyl scalar has the form 
\bea
\psi_0(x) &=& \psi_0^{(0)}+\psi_0^{(1)}+\psi_0^{(2)},
\label{eq:psiGT}\eea
with
\begin{widetext}
\bsube\bea
\psi_0^{(0)} &=& 4\pi {\mathfrak m} u^t \frac{\Delta_0^2}{r_0^2}\sum_{\ell\geq2}\sum_mA_{\ell m}[(\ell-1)\ell(\ell+1)(\ell+2)]^{1/2}R_{\rm H_{\ell,m}}(r_<)R_{\infty_{\ell,m}}(r_>){}_2Y_{\ell m}(\theta,\phi)\bar{Y}_{\ell m}\left(\frac{\pi}{2},\Omega t\right), \\
\psi_0^{(1)} &=& 8\pi i{\mathfrak m} \Omega u^t \Delta_0 \sum_{\ell\geq2}\sum_m A_{\ell m}[(\ell-1)(\ell+2)]^{1/2}
	{}_2Y_{\ell m}(\theta,\phi){}_1\bar{Y}_{\ell m}\left(\frac{\pi}{2},\Omega t\right) \times  \nonumber\\
& & \quad \Bigl\{[im\Omega r_0^2 + 2 r_0]R_{\rm H_{\ell,m}}(r_<)R_{\infty_{\ell,m}}(r_>) 
	+ \Delta_0[R_{\rm H_{\ell,m}}'(r_0)R_{\infty_{\ell,m}}(r)\theta(r-r_0) 	 + R_{\rm H_{\ell,m}}(r)R_{\infty_{\ell,m}}'(r_0)\theta(r_0-r)]\Bigr\},\\
\psi_0^{(2)} &=& -4\pi {\mathfrak m}\Omega^2 u^t \sum_{\ell\geq2}\sum_m A_{\ell m}
	{}_2Y_{\ell m}(\theta,\phi){}_2\bar{Y}_{\ell m}\left(\frac{\pi}{2},\Omega t\right) \times 
\nonumber\\ 
& & \biggl\{[30r_0^4 - 80Mr_0^3 + 48M^2r_0^2 - m^2\Omega^2 r_0^6 -2\Delta_0^2 - 24\Delta_0 r_0(r_0-M)+ 6im\Omega r_0^4(r_0-M)]
	R_{\rm H_{\ell,m}}(r_<)R_{\infty_{\ell,m}}(r_>)
 \nonumber\\ 
& & \qquad  + 2(6r_0^5 - 20Mr_0^4 + 16M^2r_0^3 - 3r_0\Delta_0^2 + im\Omega \Delta_0 r_0^4)[R_{\rm H_{\ell,m}}'(r_0)R_{\infty_{\ell,m}}(r)\theta(r-r_0) 
 + R_{\infty_{\ell,m}}'(r_0)R_{\rm H_{\ell,m}}(r)\theta(r_0-r)] \nonumber\\ 
& & \qquad + r_0^2\Delta_0^2[R_{\rm H_{\ell,m}}''(r_0)R_{\infty_{\ell,m}}(r)\theta(r-r_0) + R_{\infty_{\ell,m}}''(r_0)R_{\rm H_{\ell,m}}(r)\theta(r_0-r)+\textrm{W}[R_{\rm H_{\ell,m}}(r),R_{\infty_{\ell,m}}(r)]\delta(r-r_0)]\Biggr\}.
\eea\esube 
\end{widetext}
Here $\Delta = r^2 - 2Mr$; $\Delta_0 = r_0^2 - 2Mr_0$; $\,_sY_{\ell,m}(\theta,\phi)$ are the spin-weighted spherical harmonics; the functions $R_{\rm H_{\ell,m}}(r)$ and $R_{\infty_{\ell,m}}(r)$  are the solutions of the homogenous radial Teukolsky equation that are ingoing at the future event horizon and outgoing at null infinity, and a prime denotes a derivative with respect to $r$; the Wronskian, 
\begin{align}
\textrm{W}[R_{\rm H_{\ell,m}}(r),R_{\infty_{\ell,m}}(r)] = R_{\rm H_{\ell,m}} R_{\infty_{\ell,m}}^\prime - R_{\infty_{\ell,m}} R_{\rm H_{\ell,m}}^\prime,
\end{align}
and the quantities $A_{\ell m}$,
given by 
\be
 A_{\ell m} := \frac{1}{\Delta^3 \textrm{W}[R_{\rm H_{\ell,m}}(r),R_{\infty_{\ell,m}}(r)]},
\label{eqAlm}
\ee
are constants, independent of $r$.  

The spin-weighted spherical harmonics $_sY_{\ell,m}(\theta,\phi)$ and their $\theta$ derivatives at $\theta=\pi/2$ are calculated analytically. On the other hand, we compute the functions $R_{\rm H_{\ell,m}}$ and $R_{\infty_{\ell,m}}$ to more than 550 digits of accuracy using expansions in terms of hypergeometric functions given in Ref.~\cite{MST}, namely
\begin{widetext}
\begin{align}
R_{\rm H_{\ell,m}} &= e^{i\epsilon x}(-x)^{-2-i\epsilon}\sum_{n=-\infty}^{\infty}a_n F(n+\nu+1-i\epsilon,-n-\nu-i\epsilon,-1-2i\epsilon;x), 
\label{RH}\\
R_{\infty_{\ell,m}} &= e^{i z}z^{\nu-2} \sum_{n=-\infty}^{\infty} (-2z)^n b_n U(n+\nu+3-i\epsilon,2n+2\nu+2;-2iz),
\label{radialfunctions}
\end{align}
\end{widetext}
where $x = 1-\frac{r}{2M}$, $\epsilon = 2 M m\Omega$ and $z = -\epsilon x$. We refer the reader to \cite{MST,LivRev} for the derivation of $\nu$ (called the renormalized angular momentum), and the coefficients $a_n$ and $b_n$. Here $F$ and $U$ are the hypergeometric and the (Tricomi's) confluent hypergeometric functions. To minimize the computation time we find recurrence relations between the different $F$'s and $U$'s, so that n=0 and 1 are sufficient to find all other $n$'s in Eqs. (\ref{RH} \& \ref{radialfunctions}).

After obtaining the $\ell m$ modes of $\psi_0$, we calculate the radial part of the Hertz potential $\Psi$ using the algebraic relation
\beq
\Psi_{\ell m} = 8 \frac{(-1)^m (\ell+2)(\ell+1)\ell(\ell-1)\bar\psi_{\ell,-m}
	+ 12 i m M \Omega \psi_{\ell m} }{ [(\ell+2)(\ell+1)\ell(\ell-1)]^2 + 144 m^2 M^2 \Omega ^2},
\eeq
where the individual modes are defined as the coefficients in $\Psi = \sum_{\ell\geq2}\sum_m\Psi_{\ell m}(r){\,}_2Y_{\ell m}(\theta,\phi)e^{-im\Omega t}$ and $\psi_0 = \sum_{\ell,m}\psi_{\ell m}(r){\,}_2Y_{\ell m}(\theta,\phi)e^{-im\Omega t}$. 

From the Hertz potential, the tetrad components of the reconstructed metric perturbation are obtained using
\begin{widetext}
\bea
h^{rec}_{\bf 11} &=&  \frac{r^2}{2}(\bar{\eth}^2\Psi + \eth^2\overline\Psi), \\ 
h^{rec}_{\bf 33} &=& r^4\left[\frac{\partial_t^2 -2f\partial_t\partial_r+f^2\partial_r^2}{4} - \frac{3(r-M)}{2r^2}\partial_t
+ \frac{f(3r-2M)}{2r^2}\partial_r + \frac{r^2-2M^2}{r^4}\right]\Psi, \\
h^{rec}_{\bf 13} &=& -\frac{r^3}{2\sqrt{2}}\left(\partial_t-f\partial_r-\frac{2}{r}\right)\bar{\eth}{\Psi},
\eea
\end{widetext}
where the subscript boldfaced numbers {\bf 1} and {\bf 3} respectively represent components along $l^\alpha$ and $m^\alpha$ (the complex null vector on the 2-sphere). The other quantities appearing here are $f = \Delta/r^2$ and the operators $\eth$ and $\bar{\eth}$, the spin-raising and -lowering angular operators, which act on a spin-s quantity $\eta$ according to
\begin{eqnarray}
\eth\eta &=& -\left(\partial_\theta+i\csc\theta\partial_\phi-s\cot\theta\right)\eta, 
\label{green_eth_b}
\nonumber\\
\bar{\eth}\eta&=&-\left(\partial_\theta-i\csc\theta\partial_\phi+s\cot\theta\right)\eta.
\label{green_eth_bar_b}
\end{eqnarray}

\subsubsection{Completion}\label{completion}
As mentioned previously, the reconstructed metric perturbation is incomplete. In the context of a Schwarzchild background, the reconstructed perturbation corresponds to the $\ell>1$ tensor-spherical-harmonic modes of the solution to the linearized Einstein equation with a point particle source. To complete the solution, we must add perturbations that satisfy the remaining, $\ell=0$ and $\ell=1$, pieces of the Einstein equation. 

Appendix~\ref{appA} discusses a large range of gauge choices for the monopole solution. Here we add both monople and dipole solutions in the asymptotically flat Zerilli gauge, as given in Eqs.~(18)--(19) of Ref.~\cite{BD-spin}. The non-zero components of the $\ell=0$ mode are given by
\begin{align}
h^{\delta M}_{tt\,<} &= \frac{2\mu\hat{E}}{r_0} \frac{1-\frac{2M}{r}}{1-\frac{2M}{r_0}}, \nonumber \\
h^{\delta M}_{tt\,>} &= \frac{2\mu\hat{E}}{r}, \nonumber \\
h^{\delta M}_{rr\,<} &= 0, \nonumber \\
h^{\delta M}_{rr\,>} &= \frac{2\mu\hat{E}}{r\left( 1-\frac{2M}{r}\right)^2}, \label{hdM}
\end{align}
and those of the $\ell=1$ mode are given by
\begin{align}
h^{\delta J}_{t\phi\,<} &= - \frac{2\mu\hat{L}_z\,r^2\sin^2\theta}{r_0^3}, \nonumber \\
h^{\delta J}_{t\phi\,>} &= - \frac{2\mu\hat{L}_z\,\sin^2\theta}{r}.\label{hdJ}
\end{align}
The subscript $<$ ($>$) indicates the metric perturbation in the region $r<r_0$ ($r>r_0$). Outside the particle's orbit, at $r>r_0$, the perturbation contains invariant shifts in the spacetime's mass and angular momentum, corresponding to the contributions from the particle. Inside the particle's orbit, at $r<r_0$, the perturbations are pure gauge. We can see this explicitly by adding a gauge perturbation $\Lie_\xi g_{\mu\nu}$ with $\xi^\mu=\left(\frac{\mu\hat E}{r_0f_0}+\frac{2\mu\hat L_z}{r_0^3}\right)t^\mu \theta(r_0-r)$, which cancels the terms $h^{\delta M}_{tt\,<}$ and $h^{\delta J}_{t\phi\,<}$ given above. 

The reader may question why one choice of gauge is preferable to another, and why we cannot simply set the perturbation in the region $r<r_0$ to zero, given that we are calculating invariant quantities. The answer is that, as mentioned below Eq.~\eqref{DeltalambdaB}, the quantities are invariant only within a certain class of gauges. As is well known, the quantities we are interested in (at least as they have been formulated historically) are invariant only within the class of gauges in which the helical symmetry is manifest, in the sense that the Killing vector takes the simple form~\eqref{k}. It is also well known that these quantities are invariant only within gauges that agree on the time at infinity, which is usually resolved by working in gauges in which the asymptotic flatness is manifest, in the sense that $\lim_{r\to\infty}h_{\mu\nu}=O(1/r)$ in asymptotically Cartesian background coordinates~\cite{sbd08}.\footnote{We stress ``manifestness'' here because this is only a matter of gauge choice; the physical solution will still possess a Killing symmetry even if we switch to a gauge in which that symmetry is not plain to see, and likewise for asymptotic flatness.} However, here we point out that the quantities are invariant only within a class of gauges satisfying additional continuity conditions, which prove to be essential in our regularization scheme.

Let us first restrict to the class of gauges with manifest helical symmetry. The gauge transformations that alter our invariants even within this class are the ones which preserve the symmetry, in the sense that $\Lie_k h_{\mu\nu}=0$, but which alter the Killing vector, in the sense that $\Lie_\xi k^\mu\neq0$. To our knowledge, the most general (continuous in $r$) gauge generator that accomplishes this is given by
\beq
\xi^\mu = (\alpha_1 t+\alpha_2\phi)t^\mu + (\beta_1 t+\beta_2\phi)\phi^\mu,\label{xi-hel}
\eeq
where $\alpha_i$ and $\beta_i$ are arbitrary constants. This generates a transformation with nonzero components
\begin{align}
\delta h_{tt} &= -2 f \alpha_1,\label{dhtt}\\
\delta h_{t\phi} &= r^2\beta_1\sin^2\theta - f \alpha_2,\label{dhtp}\\
\delta h_{\phi\phi} &= 2r^2\beta_2\sin^2\theta.\label{dhpp}
\end{align}
It alters the helical Killing vector to $k^\mu\to k^\mu+\delta k^\mu$, where
\beq
\delta k^\mu = \Lie_\xi k^\mu = -(\alpha_1+\Omega\alpha_2) t^\mu-(\beta_1+\Omega\beta_2)\phi^\mu.
\eeq
By rewriting $k^\mu+\delta k^\mu$ in the form~\eqref{k} (times an overall constant), we find that this shift simply corresponds to $\Omega\to\Omega+\delta\Omega+O(\mu^2)$, with $\delta\Omega =(\alpha_1-\beta_2+\Omega\alpha_2-\beta_1/\Omega)\Omega$. 

One can reason that the transformations spoil invariance because our invariants are defined by an expansion at fixed $\Omega$, while these transformations alter $\Omega$. This makes the transformations dangerous, because the gauge perturbation \eqref{dhtt}--\eqref{dhpp} is helically symmetric, satisfying $\Lie_k\delta h_{\mu\nu}=0$. So if two people compute one of the ``invariants" in two different gauges related by this transformation, they might each think they are working in suitable gauges. But the gauge perturbation contributes to the ``invariant", and the two people will obtain different values from their calculations. To eliminate this possibility, we must impose conditions on $h_{\mu\nu}$ that fix the freedom represented by the generator~\eqref{xi-hel}. First, we notice that the $\alpha_2$ term in Eq.~\eqref{dhtp} diverges (in regular Cartesian coordinates) at $\theta=0$ and $\pi$. So we can remove that freedom by requiring regularity at the poles. Similarly, the $\beta_2$ term contains conical singularities at $\theta=0$ and $\pi$, with a deficit (or excess) angle of $2\pi\beta_2$; this is to be expected from the fact that in Eq.~\eqref{xi-hel} it corresponds to the coordinate transformation $\phi\to \phi-\beta_2\phi$.

This leaves us with the freedom in $\xi^\mu = \alpha_1 t t^\mu + \beta_1 t \phi^\mu$. At this stage, we must consider jump discontinuities at $r=r_0$. We replace $\alpha_1$ with $\alpha_1=\alpha_{1+}\theta^++\alpha_{1-}\theta^-$, and the analogue for $\beta_1$, with $\theta^\pm=\theta[\pm(r-r_0)]$. This transformation still preserves the helical symmetry of the metric perturbation, as the gauge perturbation it generates has the helically symmetric form 
\begin{align}
\delta h_{tt} &= -2 f (\alpha_{1+}\theta^++\alpha_{1-}\theta^-),\label{dhtt2}\\
\delta h_{t\phi} &= r^2(\beta_{1+}\theta^++\beta_{1-}\theta^-)\sin^2\theta. \label{dhtp2}
\end{align}
As mentioned at the beginning of this discussion, we can partly abolish this freedom by imposing asymptotic flatness. Assuming we always work in an asymptotically flat gauge, the only remaining part of the unwanted freedom is encoded in the gauge perturbations 
\begin{align}
\delta h_{tt} &= -2 f \alpha_{1-}\theta(r_0-r),\label{dhtt3}\\
\delta h_{t\phi} &= r^2\beta_{1-}\theta(r_0-r)\sin^2. \label{dhtp3}
\end{align}

How do we escape this last bit of freedom? Here we propose what we consider to be the most natural choice: we require continuity  of the $h_{tt}$ and $h_{t\phi}$ components of the completion part of the metric. This brings us back to the $\ell=0,1$ solutions in Eqs.~\eqref{hdM} and \eqref{hdJ}. We now see why we do not use the freedom in Eqs.~\eqref{dhtt3}--\eqref{dhtp3} to eliminate $h^{\delta M}_{tt\,<}$ and $h^{\delta J}_{t\phi\,<}$: doing so would violate our continuity condition.

 In summary, the ``invariant" quantities are invariant within the class of gauges satisfying conditions (i)--(iv) listed below Eq.~\eqref{DeltalambdaB}, where we have now specified the continuity condition. We stress that this continuity condition (and the regularity conditions at the poles) are not imposed in a quest for regularity. Indeed, we happily work with a perturbation that is \emph{not} in a regular gauge: The $rr$ component of Eq.~\eqref{hdM} is plainly discontinuous across $r=r_0$, and as we discuss in Appendix~\ref{appA}, this perturbation is also not regular at the horizon in horizon-penetrating background coordinates. The reconstructed perturbation is also discontinuous across $r=r_0$, a fact which we find experimentally and discuss further in the next section. But none of these gauge irregularities affect the values of the invariants, because they are not generated by transformations of the form~\eqref{xi-hel}. 

Since the ``invariants" do not have unique values, in principle we could equally well choose some other, less regular class of gauges to work in. Here we take a pragmatic stance: we want our invariants to agree with those that would be calculated by anyone working in a ``nice" gauge that is asymptotically flat and continuous everywhere away from the particle (at least for $r>2M$, where the background coordinates are well defined). Clearly such a ``nice" gauge falls within the class defined by our four conditions.

Although we have specified the continuity condition at $r=r_0$, properly speaking it must be imposed at all radii. But it becomes important only at $r=r_0$, because only there is one tempted to introduce a discontinuity. Especially in Kerr, setting the perturbation to zero in the region $r<r_0$ often appears the easy choice to make. It is doubly tempting because the perturbation is often already discontinuous across $r=r_0$, as is the case for us here. 

If one is able to calculate the invariants from modes in the region $r>r_0$, one need not worry about the issue of continuity. However, our regularization scheme will use the average of modes from $r\to r_0^+$ and $r\to r_0^-$. Hence, to ensure that our invariants have the same values as would be found in a ``nice" gauge, we must impose the continuity conditions at $r=r_0$.


\section{Regularization}\label{sec3}
From the retarded field we now wish to calculate the  invariants $\Delta\psi$, $\Delta\lambda_i^{\rm E}$, and $\Delta\lambda^{\rm B}$. In Sec.~\ref{sec2}, we defined these invariants in terms of the DW regular field, which was originally defined in the Lorenz gauge~\cite{Detweiler:2002mi}. While it is easily generalized to any gauge smoothly related to Lorenz, it is less easily generalized to more singular gauges, among them the radiation gauge. In this section, we utilize the results of PMB to define a DW regular field in a completed radiation gauge; we show that the invariants constructed from this field are equal to those constructed from the regular field in any gauge smoothly related to Lorenz. Motivated by the parity structure of the transformation between the Lorenz gauge and radiation gauge, we compute the invariants using a mode-sum regularization formula that averages the $\ell$ modes from opposite sides of the particle. The validity of this procedure will be established in a companion paper.

\subsection{Detweiler-Whiting fields}\label{DW fields}

Radiation-gauge metric perturbations are known to have singularities extending away from the particle, either in the form of a string emanating from the particle or in the form of distributional singularities supported on a surface intersecting the particle~\cite{Pound-Merlin-Barack:14}. In our calculations in this paper, we work in the ``no-string" gauge described by PMB; in the present context, this gauge is regular everywhere except at the sphere at $r=r_0$, which supports a jump discontinuity and a delta function. 

Rather than working directly with the metric perturbation, for our analysis we will instead refer to the gauge transformation from any gauge smoothly related to Lorenz. Let $x_0^\mu$ be a point on the zeroth-order worldline $\gamma_0$. At a nearby point $x^\mu=x_0^\mu+\delta x^\mu$, the generator of the gauge transformation has the discontinuous form
\begin{align}
\xi^\mu &= [\xi^\mu_{0+}(x_0,\delta x)+Z^\mu_+(x_0)+O(\varepsilon\ln\varepsilon)]\theta^+\nonumber\\
				& +[\xi^\mu_{0-}(x_0,\delta x)+Z^\mu_-(x_0)+O(\varepsilon\ln\varepsilon)]\theta^-.
\end{align}
Here $\varepsilon$ is a measure of distance from $x_0^\mu$, such that $\delta x^\mu\sim\varepsilon$. The quantities $Z^\mu_\pm(x_0)$ are smooth functions of $x_0^\mu$. The other terms, $\xi^\mu_{0\pm}$ are irregular at the particle. It is convenient to split them into pieces that become parallel and perpendicular to the worldline as $\varepsilon\to0$:
\beq
\xi^\mu_{0\pm} = \xi_{\parallel\pm}(x_0,\delta x)u^{\mu}(x_0)+\xi^\mu_{\perp\pm}(x_0,\delta x),
\eeq
where $u_\mu(x_0)\xi^\mu_{\perp\pm}(x_0,\delta x)=0$. The parallel piece behaves as
\beq
\xi_{\parallel\pm}(x_0,\delta x)\sim\ln(\varepsilon).\label{xipar}
\eeq
Each of the perpendicular pieces $\xi^\mu_{\perp+}$ and $\xi^\mu_{\perp-}$ is bounded but has a direction-dependent limit $\delta x^\mu\to0$. The two are related by
\beq 
\xi^\mu_{\perp+}(x_0,-\delta x)=-\xi^\mu_{\perp-}(x_0,\delta x);\label{xiperp}
\eeq
in other words, $\xi^\mu_{\perp}:=\xi^\mu_{\perp+}\theta^++\xi^\mu_{\perp-}\theta^-$ has odd parity under the parity transformation that interchanges points on diametrically opposite sides of the particle. Its derivatives $\partial_\alpha\xi_{\beta\perp}$ have even parity; this is also true of $\partial_\alpha\xi_{\parallel}u_\beta$, even though $\xi_\parallel$ has no definite parity. These general properties, derived by PMB, are all that we will require. We refer the reader to Ref.~\cite{Pound-Merlin-Barack:14} for explicit expressions. Given our construction of the metric, and the helical symmetry, continuity, and asymptotic flatness of the (tweaked-for-asymptotic-flatness~\cite{sbd08}) Lorenz-gauge solution, it is clear that $\xi^\mu$ satisfies the condition $\Lie_k\xi^\mu=0$.

The retarded field in the no-string gauge is given by
\beq\label{h-transform}
h_{\mu\nu} = h^L_{\mu\nu} + \Lie_\xi g_{\mu\nu},
\eeq
where we will use ``L" to indicate that a quantity is in the Lorenz gauge. (An arbitrary gauge that is smoothly related to Lorenz could equally well be used.) As promised, this gauge is discontinuous across the sphere at $r=r_0$, due to the $\theta^\pm$ functions in $\xi^\mu$, and it contains a delta function on that sphere, due to derivatives of those functions. Our reconstruction and completion procedure cannot determine the correct $\delta$ function on the sphere to make our computed metric perturbation a true solution to the linearized Einstein equation with a point-particle source, but PMB showed that the perturbation we obtain is pointwise correct away from $r=r_0$. Hence, we shall always work away from the sphere.

PMB established that in this irregular gauge, the linear correction $\delta z^\mu$ to the particle's trajectory, relative to the trajectory $z^\mu$ in the Lorenz gauge, is given by the average of the regular part of the gauge transformation:
\beq\label{dz}
\delta z^\mu = -\frac{1}{2}(Z^\mu_++Z^\mu_-).
\eeq
This is the regular change in position induced by the singular generator $\xi^\mu$.

How can we use this to define a Detweiler-Whiting regular field in the no-string gauge? The DW regular field possesses three important properties: (i) the linear equation of motion can be written as the geodesic equation in $g_{\mu\nu}+h^R_{\mu\nu}$, (ii) $h^R_{\mu\nu}$ is a smooth vacuum solution to the linearized Einstein equation, and (iii) when evaluated at a point $x_0^\mu$ on the worldline, $h^R_{\mu\nu}$ and all its derivatives depend only on the causal past of $x^\mu_0$. To simplify the analysis, let us ignore the third property and focus on the first two. Consider a smooth transformation away from Lorenz. It corresponds to a coordinate transformation $x^\mu\to x^\mu-\xi^\mu$, altering the position of the particle to $z^\mu+\delta z^\mu$ with $\delta z^\mu=-\xi^\mu(x_0)$. It also alters the metric perturbation to $h^L_{\mu\nu}+\Lie_\xi g_{\mu\nu}$. A priori, there is no unique way of splitting $h^L_{\mu\nu}+\Lie_\xi g_{\mu\nu}$ into singular and regular pieces. However, if for a moment we ignore the singular field and consider a geodesic in a vacuum metric $g_{\mu\nu}+h^R_{\mu\nu}$, it is obvious that since the Einstein equation and geodesic equation are invariant, $h^R_{\mu\nu}+\Lie_{\xi}g_{\mu\nu}$ will remain a vacuum perturbation, and $z^\mu+\delta z^\mu$ will be a geodesic in $g_{\mu\nu}+h^R_{\mu\nu}+\Lie_{\xi}g_{\mu\nu}$. This motivates the natural choice of singular and regular fields in the new gauge:
\begin{align}
h^R_{\mu\nu} &= h^{RL}_{\mu\nu}+\Lie_{\xi}g_{\mu\nu},\\
h^S_{\mu\nu} &= h^{SL}_{\mu\nu}.
\end{align}
With this choice, $h^R_{\mu\nu}$ is guaranteed to satisfy properties (i) and (ii) mentioned above.

Now let us consider our singular gauge transformation to the no-string gauge. Suppose we split $\xi^\mu$ into regular and singular pieces of the form
\begin{align}
\xi^\mu_R &= \frac{1}{2}\left(Z^\mu_++Z^\mu_-\right)+O(\varepsilon),\label{xiR}\\
\xi^\mu_S &= \xi^\mu-\xi^\mu_R,
\end{align}
where the $O(\varepsilon)$ terms are smooth but for our purposes can be arbitrary. Then according to Eq.~\eqref{dz} we have $\delta z^\mu = -\xi^\mu_R(x_0)$. Hence, from the argument laid out in the case of a smooth transformation, the regular field
\beq\label{hRdef}
h^R_{\mu\nu} = h^{RL}_{\mu\nu}+\Lie_{\xi_R}g_{\mu\nu}
\eeq
will satisfy the desired properties (i) and (ii): the perturbed motion is geodesic in $g_{\mu\nu}+h^R_{\mu\nu}$, and $h^R_{\mu\nu}$ is a smooth vacuum perturbation. With this choice, the singular field is left to be
\begin{align}\label{hSdef}
h^S_{\mu\nu} &= h^{SL}_{\mu\nu}+\Lie_{\xi_S}g_{\mu\nu}.
\end{align}
These will be our definitions of the singular and regular fields in the no-string gauge.

\subsection{Invariance of the invariants}
We now consider the construction of scalar quantities from $h^R_{\mu\nu}$ (and its derivatives) on the worldline. Let $\Delta I$ be one such quantity. It can be any one of Detweiler's redshift, $\Delta\psi$, $\Delta \lambda^{\rm E}_i$, $\Delta \lambda^{\rm B}$, etc., defined by substituting the no-string-gauge $h^R_{\mu\nu}$ into Eqs.~\eqref{Deltapsi}--\eqref{DeltalambdaB}. Before describing the computation of this quantity, we first confirm that it is equal to its value $\Delta I^L$ in the Lorenz gauge (and hence to its value in any gauge smoothly related to Lorenz).

Let us begin by defining retarded and singular versions of $\Delta I$. We do this by extending $\Delta I$ away from the worldline; for example, in Eqs. \eqref{Deltapsi}--\eqref{DeltalambdaB} we can multiply each term by an arbitrary function of $t,r,\theta,\phi$ that goes smoothly to 1 at $x_0$, and at points off $x_0$, we can let $h^R_{\mu\nu}$ and its partial derivatives take their natural values. This defines a field $\Delta \hat I[h^R]$. If we then replace the regular field with the retarded or singular field, we obtain fields $\Delta \hat I[h]$ and $\Delta \hat I[h^S]$.

$\Delta I$ can be written as the limit of the difference between these two fields,
\beq\label{lim}
\Delta I (x_0) = \lim_{x\to x_0}\left\{\Delta \hat I[h](x)-\Delta \hat I[h^S](x)\right\}.
\eeq
Note that while the extension is arbitrary, it must be the same in both terms on the right. Starting from Eq.~\eqref{lim}, we can establish our desired result, $\Delta I=\Delta I^L$, by transforming to the Lorenz gauge before taking the limit. In the extended fields $\Delta \hat I$, the only gauge-dependent quantities are the perturbations $h_{\mu\nu}$, $h^R_{\mu\nu}$, and $h^S_{\mu\nu}$; the only other dependencies are on the zeroth-order trajectory, which is gauge-invariant~\cite{Pound:14c}, and on arbitrary smooth functions that define the extension, which we wish to keep the same in both gauges. Hence, the quantities constructed from no-string fields are related to quantities constructed from Lorenz-gauge fields according to
\begin{align}
\Delta \hat I[h] &= \Delta\hat I[h^L]+\Delta\hat I[\Lie_{\xi_R} g_{\mu\nu}]+\Delta\hat I[\Lie_{\xi_S} g_{\mu\nu}],\\
\Delta \hat I[h^S] &= \Delta\hat I[h^{SL}] + \Delta\hat I[\Lie_{\xi_S} g_{\mu\nu}],
\end{align}
where we have appealed to Eqs.~\eqref{h-transform} and \eqref{hSdef}. Since $\Delta I$ is invariant under gauge transformations generated by smooth vectors $\xi^\mu$ satisfying $\Lie_k\xi^\mu=0$, we have $\lim_{x\to x_0}\Delta\hat I[\Lie_{\xi_R} g_{\mu\nu}]=0$. The $\xi^\mu_S$ terms in Eq.~\eqref{lim} cancel, leaving us with
\beq
\Delta I (x_0) = \lim_{x\to x_0}\left\{\Delta \hat I[h^L](x)-\Delta \hat I[h^{SL}](x)\right\},
\eeq
which is our desired result:
\beq
\Delta I = \Delta I^L.
\eeq

We have now shown that the invariants constructed from the no-string-gauge regular field are equal to the ones constructed from the usual DW field in gauges smoothly related to Lorenz. However, we note that we could have split $\xi^\mu$ into $\xi^\mu_R$ and $\xi_S^\mu$ in any number of other ways and still obtained the same result. For example, we could have defined $\xi^\mu_R=0$, and the calculations in this section would have gone through just as well. In that case, the regular field in the no-string gauge would have been defined to be identically equal to the Lorenz-gauge regular field. The message is that for the purposes of calculating gauge-invariant scalars on the worldline, in Eqs.~\eqref{hRdef} and \eqref{hSdef} the regular gauge vector $\xi_R^\mu$ is almost entirely arbitrary. So long as $\xi_R^\mu$ is smooth and helically symmetric, the choice has no effect on the invariants. Nevertheless, we think it is meaningful to define the regular metric to be the one in which the perturbed motion is geodesic.

\subsection{Averaged mode-sum regularization}
Assured that we are calculating the correct quantities $\Delta I$, we now describe our concrete method of performing that calculation. PMB showed that the gravitational self-force in the no-string gauge can be computed using mode-sum regularization in combination with averaging from opposite sides of the particle. Here we adopt the same method for our calculation of invariants.

First note that the direction of the limit in Eq.~\eqref{lim} is arbitrary so long as it avoids the sphere at $r=r_0$. Therefore we can replace the limit with the average of two limits from opposite directions,
\begin{align}
\Delta I &= \lim_{\delta x\to 0}\left[\langle\Delta \hat I[h]\rangle-\langle\Delta \hat I[h^{SL}]\rangle\right],
\end{align}
where $x_0+\delta x$ lies outside the sphere and $x_0-\delta x$ lies inside it, and to keep the expressions compact, we have defined the average of a field $F$ as $\langle F \rangle(\delta x) = \frac{1}{2}\left[F(x_0+\delta x)+F(x_0-\delta x)\right]$.

Since we have a great deal of knowledge about the Lorenz-gauge singular field, we express $h^S_{\mu\nu}$ in terms of it, giving us
\begin{align}
\Delta I &= \lim_{\delta x\to 0}\left[\langle\Delta \hat I[h]\rangle-\langle\Delta \hat I[h^{SL}]\rangle
						-\langle\Delta \hat I[\Lie_{\xi_S}g]\rangle\right].\label{<DI>}
\end{align}
Following standard steps~\cite{Barack-Ori:00}, we can rewrite this as a sum over scalar spherical harmonic modes, 
\begin{align}\label{modesum1}
\Delta I &= \frac{1}{2}\sum_{\ell=0}^\infty\sum_{k=\pm}\left[(\Delta \hat I[h])_k^{\ell}-(\Delta \hat I[h^{SL}])_k^{\ell}
						-(\Delta \hat I[\Lie_{\xi_S}g])_k^{\ell}\right],
\end{align}
where $(\Delta \hat I)^{\ell}(t,r,\theta,\phi)$ is the $\ell$th term in the spherical harmonic expansion of $\Delta \hat I$, summed over azimuthal number $m$, and $(\Delta \hat I)_\pm^{\ell}=\lim_{r\to r_0^\pm}(\Delta \hat I)^{\ell}(t,\delta r,\pi/2,\Omega t)$ is its limit to the particle from $r>r_0$ or $r<r_0$. (Because the orbit is circular, this quantity depends only on $r_0$ in the end.) 

For concreteness, let us examine $\Delta\hat\psi$. We note that the singular field in the Lorenz gauge behaves as $h^{SL}_{\mu\nu}\sim 1/\varepsilon$. Since $\Delta\hat\psi^{SL}$ involves single derivatives of $h^{SL}_{\mu\nu}$, it behaves as $\Delta \hat\psi[h^{SL}]\sim 1/\varepsilon^2$. It is well known that this translates into the large-$\ell$ form 
\begin{align}
(\Delta \hat\psi[h^{SL}])_\pm^{\ell} &= A^{\psi L}_{1\pm} L + A^{\psi L}_{0\pm} +A^{\psi L}_{-1\pm}/L + O(1/L^2),
\end{align}
where $L:=\ell+1/2$ and the regularization parameters $A^{\psi L}_{n\pm}$ are independent of $\ell$. From the general property $\xi^\mu\sim\ln\varepsilon$, we expect the same scalings for $(\Delta \hat\psi[\Lie_{\xi_S}g])_\pm^{\ell}$, and we assume 
\begin{align}\label{dDpsiell}
(\Delta \hat\psi[\Lie_{\xi_S}g])_\pm^{\ell} &= \delta A^\psi_{1\pm} L + \delta A^\psi_{0\pm} + \delta A^\psi_{-1\pm}/L + O(1/L^2).
\end{align}
Based on this, we write Eq.~\eqref{modesum1} as
\begin{align}
\Delta \psi &= \frac{1}{2}\sum_{\ell,k=\pm}\left[(\Delta \hat\psi[h])_k^{\ell}-A^\psi_{1k}L - A^\psi_{0k} - A^\psi_{-1k}/L \right]\nonumber\\
&\quad-\frac{1}{2}(D^\psi_+ + D^\psi_-),\label{modesumpsi}
\end{align}
where $A^\psi_{n\pm}=A^{\psi L}_{n\pm}+\delta A^{\psi}_{n\pm}$, and
the $D$ terms contain everything in the singular field modes that is not included in the $A$ terms. Explicitly, $D^\psi_\pm=D^{\psi L}_\pm+\delta D^\psi_{\pm}$, 
\begin{align}
D^{\psi L}_\pm &:= \sum_{\ell=0}^\infty[(\Delta\hat \psi[h^{SL}])^{\ell}_\pm -A^{\psi L}_{1\pm} L - A^\psi_{0\pm} - A^\psi_{-1\pm}/L],\!
\end{align}
and the analog for $\delta D^{\psi}_\pm$. Note that the parameters $A^{\psi}_{n\pm}$ and $D^\psi_\pm$ depend on the choice of extension in defining the extended field $\Delta \hat\psi[h^S]$, and they must all be calculated in the same extension as is $(\Delta\hat \psi[h])^{\ell}_\pm$. 


Using the method developed in Refs.~\cite{sf2,sf3,sf4}, we can numerically (and uniquely) determine the $A_{i\pm}$ parameters through large-$\ell$ fits of the modes of the retarded field. This allows us to numerically confirm the expected behavior~\eqref{dDpsiell}. However, there is no obvious way of determining $D^\psi_\pm$ through numerical fits, and these $D$ terms become the central concern. Here we eliminate that concern by contending that in the averaged form~\eqref{modesumpsi}, the $D$ terms cancel in the mode-sum formula, meaning they need not be determined.

First we consider the contribution to $D^\psi_\pm$ from $h^{SL}_{\mu\nu}$. We argue that the parameters $D^{\psi L}_\pm$ always vanish, regardless of the choice of extension. To see this, observe that $h^{SL}_{\mu\nu}$ has the schematic form~\cite{Barack:09}
\beq\label{hS-form}
h^{SL}_{\mu\nu} \sim \frac{1}{\rho}+\frac{(\delta x)^3}{\rho^3} + O(\varepsilon),
\eeq
where $\rho\sim\varepsilon$ is the leading-order geodesic distance from $x_0^\mu$ to $x_0^\mu+\delta x^\mu$. Any quantity constructed from $h^S_{\mu\nu}$ via single derivatives and multiplication by smooth functions will have the form
\beq\label{dhS-form}
s(x)\partial h^{SL}_{\mu\nu} \sim \frac{\delta x}{\rho^3}+\frac{(\delta x)^3}{\rho^5}+\frac{(\delta x)^7}{\rho^7} + O(\varepsilon),
\eeq
where $s(x)$ is smooth but arbitrary. Similarly, for two derivatives, as appear in the tidal invariants,
\beq\label{ddhS-form}
s(x)\partial^2 h^{SL}_{\mu\nu} \sim \frac{(\delta x)^2}{\rho^5}+\frac{(\delta x)^5}{\rho^7}+\frac{(\delta x)^8}{\rho^9}+\frac{(\delta x)^{11}}{\rho^{11}} + O(\varepsilon).
\eeq
Now, since they are finite and hence can contain no positive powers of $L$ in their mode decomposition, the $O(\varepsilon^0)$ terms may seem likely to generate a $D$ term. However, these $O(\varepsilon^0)$ terms always have the discontinuous (direction-dependent) form $(\delta x)^n/\rho^n$, with odd $n$. The sum of modes of a discontinuous quantity evaluated at the point of discontinuity, $\sum_\ell(\delta x)^n/\rho^n$ in this case, converges to the average of the quantity on an infinitesimal circle around the point~\cite{Sansone:77}. Because $(\delta x)^n/\rho^n$ terms with odd $n$ have odd parity around the point, they contribute nothing to the mode sum, and hence contribute no $D$. But the remaining terms in Eqs.~\eqref{hS-form}--\eqref{ddhS-form} (that do not vanish in the limit $\varepsilon\to0$) only ever contribute non-negative powers of $L$ to the mode sum~\cite{Leor,Barry}; in fact, this should be true of any functions of the form $(\delta x)^n/\rho^p$ with odd $p>0$~\cite{Barry}. So they too cannot produce a $D$. 

This argument should be easy to make precise and to extend to quantities constructed from any number of derivatives. Hence, we conclude that $D$ terms vanish identically for any quantity constructed from the Lorenz-gauge singular field via the action of derivatives and multiplication by smooth functions. This conclusion is independent of extension, since a change in extension merely corresponds to a change in the smooth functions multiplying the derivatves. Therefore, we have $D^{\psi L}_\pm = 0$.

Now let us turn to $\delta D^{\psi}_\pm$. These $D$ terms cancel in Eq.~\eqref{modesumpsi} if they have the symmetry property $\delta D^\psi_-=-\delta D^\psi_+$. Establishing this property requires a detailed local analysis of $\Delta \hat\psi[\Lie_{\xi_S}g]$, which will be carried out in a companion paper. In the meantime, we present a plausibility argument. From the general properties of $\xi^\mu$ described in Sec.~\ref{DW fields}, it follows that $\Delta \hat\psi[\Lie_{\xi_S}g]$ has the form
\begin{align}\label{Dpsi-eps}
\Delta \hat\psi[\Lie_{\xi_S}g]&\sim [1+\delta x+(\delta x)^2]\partial^2\xi_{S}+[1+\delta x]\partial \xi_S \nonumber\\
		&\quad + \xi_S + o(\varepsilon^0)
\end{align}
where ``$o(\varepsilon^0)$'' means ``goes to zero in the limit $\varepsilon\to0$''. Now let us assume that all positive powers of $L$ in $(\Delta \hat\psi[\Lie_{\xi_S}g])_\pm^\ell$ arise from the negative powers of $\varepsilon$ in Eq.~\eqref{Dpsi-eps}, and that $\delta D^\psi_\pm$ arises from the order-$\varepsilon^0$ terms. For the purpose of this sketch, consider only the most singular terms in $\partial\xi_S^\mu$. Given this restriction, order-$\varepsilon^0$ terms in Eq.~\eqref{Dpsi-eps} can only arise in the forms $(\delta x)^2\partial^2\xi_{S}$, $\delta x\partial \xi_S$, and $\xi_S$. Recall that the most singular terms in $\partial\xi_S$ have even parity, meaning $\delta x\partial \xi_S$ has odd parity; hence, this term will vanish upon averaging. Similarly, the most singular term in $\partial^2\xi_{S}$ has odd parity, and so $(\delta x)^2\partial^2\xi_{S}$ also vanishes upon averaging. Last, the $\xi_S$ term in Eq.~\eqref{Dpsi-eps} can only come in the form $\xi_S^\mu\partial_\mu g_{\alpha\beta}$, stemming from such a term in $\Lie_{\xi_S}g_{\alpha\beta}$. Since $u^\mu\partial_\mu g_{\alpha\beta}=0$, only $\xi_\perp^\mu$ contributes to this term; and since $\xi_\perp^\mu$ has odd parity, this contribution also vanishes under averaging. Therefore, if $\delta D^\psi_\pm$ is generated solely from the terms in Eq.~\eqref{Dpsi-eps} that have finite, nonzero limits as $\varepsilon\to0$, then averaging eliminates the $D$ terms in Eq.~\eqref{modesumpsi}, as desired.

Following these arguments, we now allow ourselves to set $D_\pm$ to zero in Eq.~\eqref{modesumpsi}. This gives us our final formula for $\Delta \psi$:
\begin{align}
\Delta \psi &= \frac{1}{2}\sum_{\ell,k=\pm}\left[(\Delta\hat \psi[h])^{\ell}_k-A^\psi_{1k}L - A^\psi_{0k}\right].\label{psi-modesum-final}
\end{align}
Here we have used $A^\psi_{-1\pm}=0$, a fact which we find numerically. 

Following identical arguments for the tidal invariants, we obtain
\begin{align}
\Delta \lambda &= \frac{1}{2}\sum_{\ell,k=\pm}\left[(\Delta\hat \lambda[h])^{\ell}_k-A^\lambda_{2k}L^2-A^\lambda_{1k}L- A^\lambda_{0k}\right],\label{lambda-modesum-final}
\end{align}
where $\Delta \lambda$ is any of $\Delta \lambda^{\rm E}_i$ or $\Delta\lambda^{\rm B}$.

When implementing these formulas, we first determine the $A$ parameters numerically. We find that the parameters, with our choice of extension, agree precisely with the analytically derived parameters in the Lorenz gauge in Ref.~\cite{Dolanetal-tidal}, with the lone exception of $A^\lambda_{0k}$. 
After making this determination, in practice we use the analytical parameters except in the case of $A^\lambda_{0k}$, which we determine numerically.

As an aside, we note that one need not always define the modes in these sums as scalar spherical harmonic modes. It is computationally simpler to leave the harmonics in the ``natural'' form they arrive in from the reconstruction procedure, which is a mixture of harmonics of various spin weights; using this ``natural'' form avoids having to re-expand spin-weighted harmonics into scalar harmonics. Although strictly speaking the arguments made in this section assume the modes are defined in the scalar harmonic basis, we have experimentally found that our results are unaltered by changing between the ``natural'' and scalar basis. It is has been suggested to us that this may hinge on our averaging from two sides of the particle~\cite{Barry}, as such averaging is known to annihilate $D$ terms that arise in some cases when nonscalar bases are used. However, we leave exploration of this issue to future work.


\section{Extraction of pN coefficients}
\label{sec4}

In our concrete implementation of the mode-sum formulas~\eqref{psi-modesum-final}--\eqref{lambda-modesum-final}, we follow the method of Ref.~\cite{sf3}. We fit the $\ell$ modes to polynomials in $L$ to numerically determine the regularization parameters $A_{i\pm}$, and to gain faster convergence, we also fit the power-law ``tail" of the singular field, which is made up of negative powers of $L$. 

Because we wish to extract pN coefficients, we calculate the invariants at very large orbital radii. Specifically, we calculate the invariants for 144 different values of $\Omega$ by placing the particle at 
\bea \label{rsvalues}
R = 1,2,3,...,9\times 10^{18,19,20,...,33},
\eea
where $R = (M\Omega)^{-2/3}$ is a coordinate-independent measure of orbital radius, which in Schwarzschild coordinates is equal to the orbital radius $r_0$ of our zeroth-order trajectory. To obtain our desired accuracies of 1 part in $10^{500}$, we calculate $\ell$ modes up to a maximum $\ell_{max}=200$. These extreme accuracies are made possible with Mathematica's arbitrary-precision algebra.

To extract the pN coefficients from these results, we fit the high-precision numerical data to the pN series in the following manner: First, we utilze the known pN series calculated by Bini and Damour, which were obtained to 8.5pN order for $\delta\psi$ and 7.5pN order for the tidal invariants. We subtract these series from the numerical data and then fit the result to a pN series $y^j\log^i(y)$, where $y=1/R$; $i/2$ is the pN order of a term. In performing the fit we first extract as many analytical coefficients of the highest power of $\log(y)$ for a given pN order as possible. This is done before anything else because the analytical coefficient multiplying the highest power of log is always either a rational number (for integer powers of $y$ in $\delta\psi$ and $\Delta\lambda_n^E$, and for half-integer powers of $y$ in $\Delta\lambda^B$) or a rational number times $\pi$ (for half-integer powers of $y$ in $\delta\psi$ and $\Delta\lambda_n^E$, and for integer powers of $y$ in $\Delta\lambda^B$). Extracting two whole numbers ($p$ and $q$ of $p/q$ or $p\pi/q$) from the numerical value requires less precision than extracting rational coefficients of transcendentals like $\pi^2$, $\gamma$, $\log(2)$, $\log(3)$, etc., which make up the pN coefficients of less than maximum power of $\log(y)$ for a given pN order. Once the coefficients of the highest power of $\log(y)$ are calculated we subtract them (along with the known pN series) from the numerical data and fit the resultant to the pN series (without the terms whose coefficients are known) and extract the coefficients of the second highest power of $\log(y)$. These coefficients usually have $\pi^2$, $\gamma$, $\log(2)$ and $\log(3)$ in them. We repeat these two steps until as many analytical coefficients are extracted as possible. The third step is to subtract all the analytical coefficients from the data and fit the remaining pN series, the coefficients in which are determined as numbers with finite accuracy rather than as analytical expressions. This procedure, explained in more detail in Ref.~\cite{Nathanetal}, enables us to obtain the pN expansion to significantly higher order than if we performed a straightforward numerical fit of the data to the series $y^j\log^i(y)$.

\begin{figure*} [htb]
\centering
\includegraphics[width=0.65\textwidth]{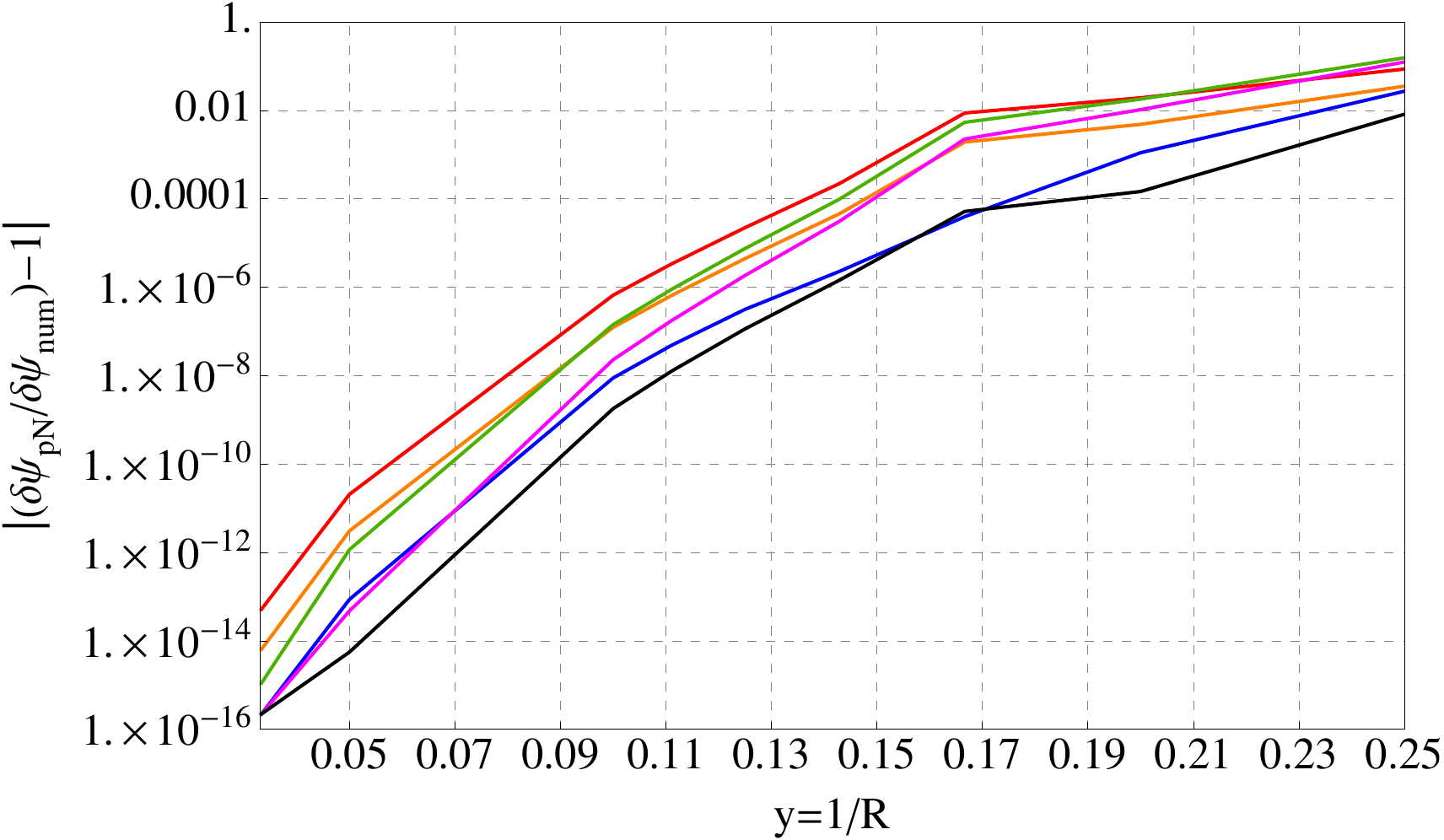}
\caption{Fractional difference between the  different pN-order-approximations and the numerical value of $\delta\psi$. Color-code is as follows - Black: 19pN, Magenta: 18pN, Blue: 17pN, Green: 16pN , Orange: 15pN, Red: 14pN.}
\label{fig1}
\end{figure*}

\begin{figure*} [htb]
\centering
\includegraphics[width=0.495\textwidth]{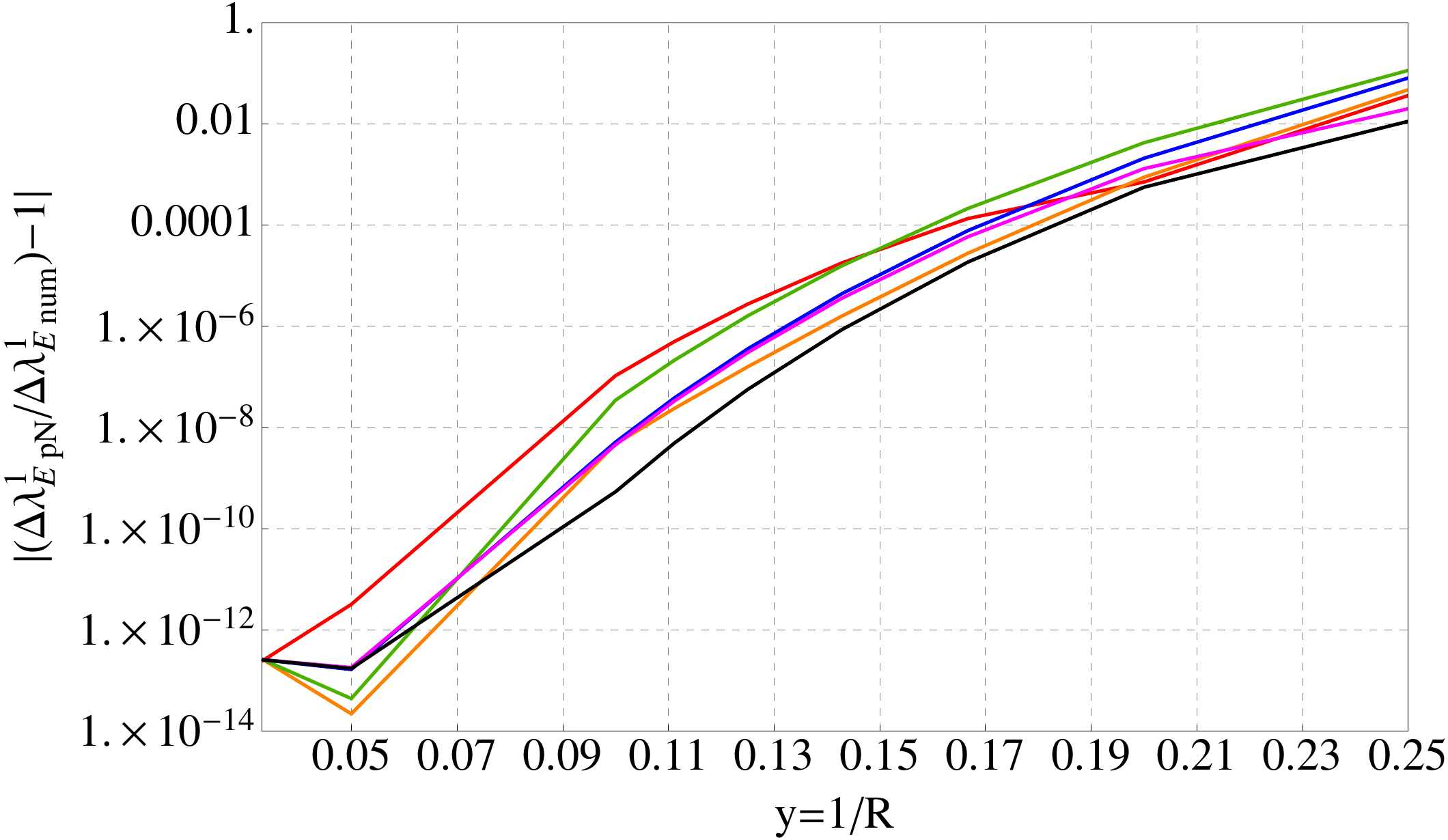}
\includegraphics[width=0.495\textwidth]{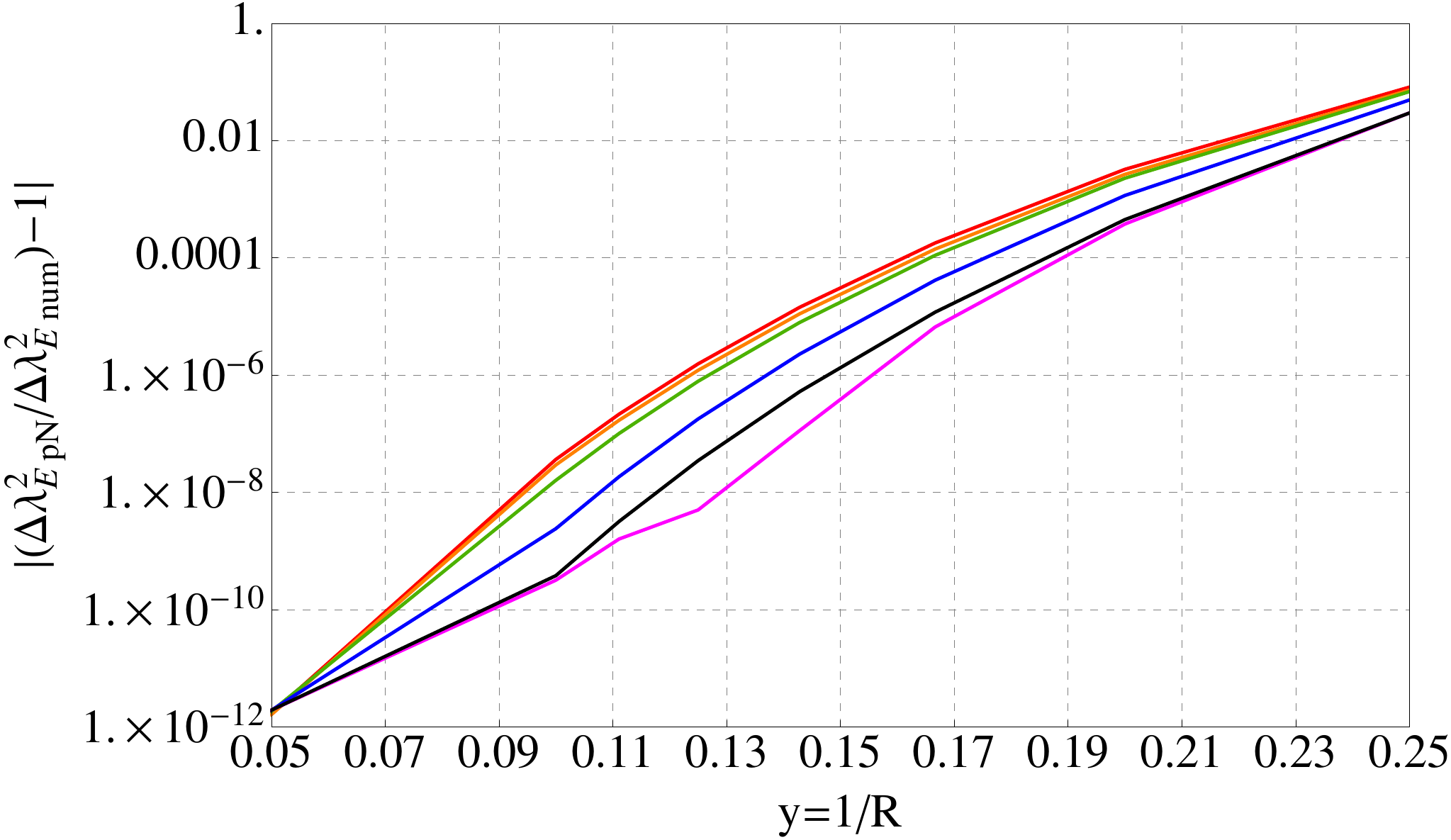}
\includegraphics[width=0.495\textwidth]{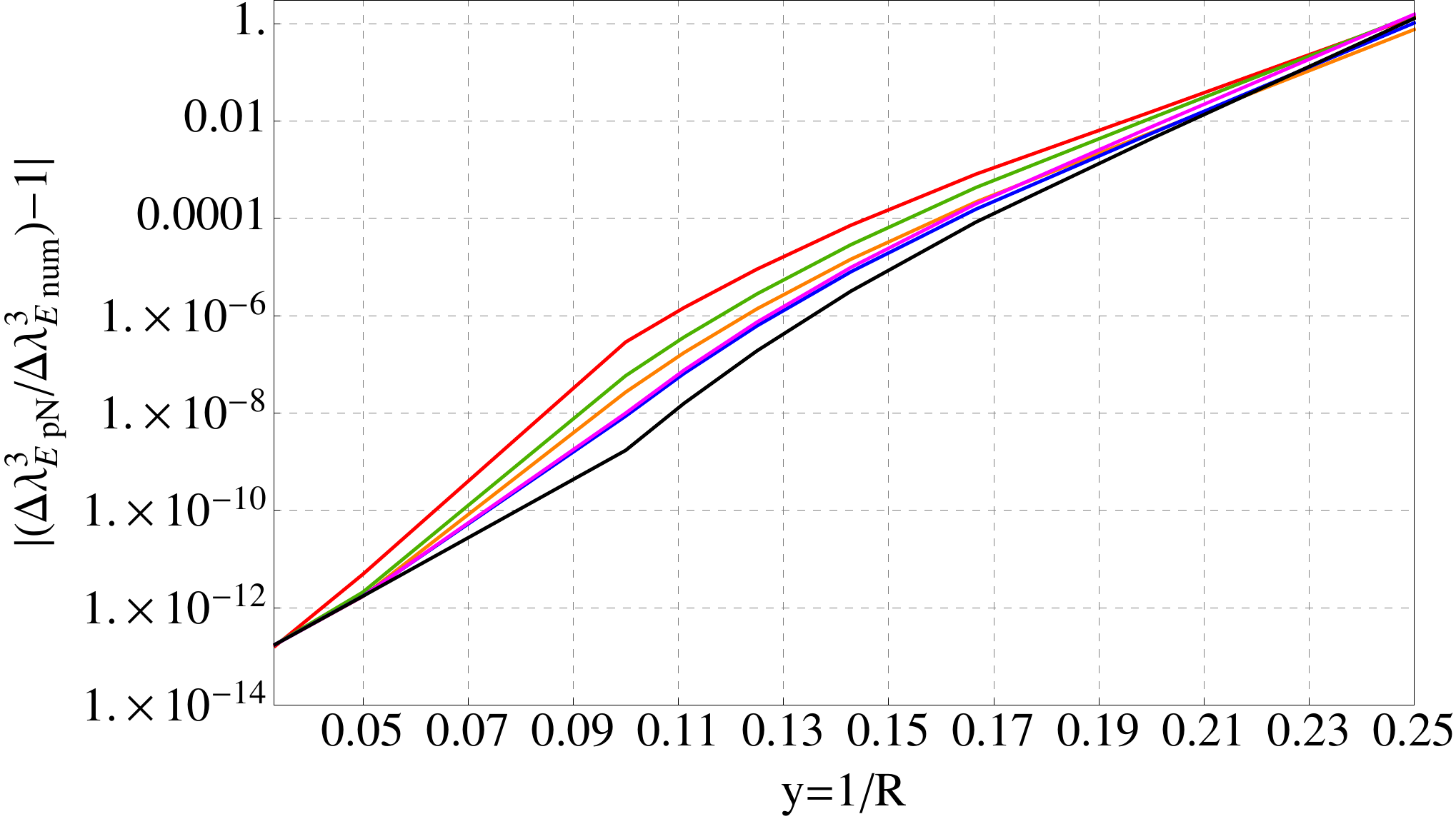}
\caption{Fractional difference between the  different pN-order-approximations and the numerical values of $\Delta\lambda_1^E$, $\Delta\lambda_2^E$, and $\Delta\lambda_3^E$. Color-code is as follows - Black: 20pN, Magenta: 19pN, Blue: 18pN, Green: 17pN , Orange: 16pN, Red: 15pN.}
\label{fig2}
\end{figure*}

\begin{figure*} 
\centering
\includegraphics[width=0.65\textwidth]{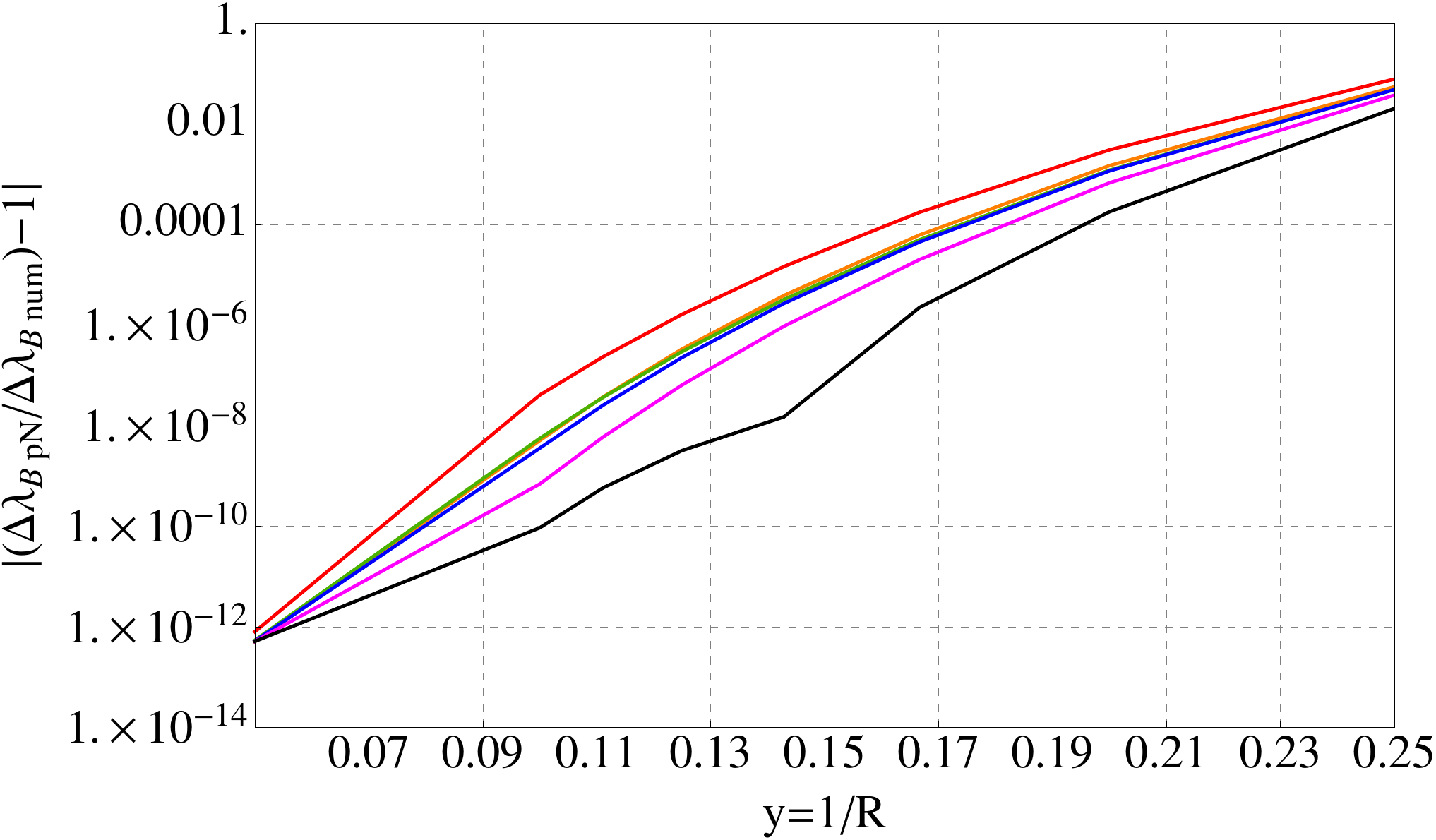}
\caption{Fractional difference between the  different pN-order-approximations and the numerical value of $\Delta\lambda^B$. Color-code is as follows - Black: 20.5pN, Magenta: 19.5pN, Blue: 18.5pN, Green: 17.5pN , Orange: 16.5pN, Red: 15.5pN.}
\label{fig5}
\end{figure*}

In Figs.~(\ref{fig1}$-$\ref{fig5}), we compare the pN-approximated invariants with their exact (up to numerical error) values computed in Ref.~\cite{Dolanetal-tidal}, and we find good agreement all the way to near $R=1/y=4M$. More precisely, we find that our pN expressions are accurate to more than 4 digits up to the innermost stable circular orbit ($R=6M$), and the approximation then worsens to less than 1 digit at $R=4M$. We also examine the efficacy of exponentially resumming the pN series (as described in Ref.~\cite{Isoyamaetal}). Figs.~(\ref{fig6}-\ref{fig8}) show the relative errors in the pN series and in the exponentially resummed series for the tidal invariants at $R=4,\,5,\,6M$. We see that resummation does not markedly improve the accuracy of the series. As a consistency check, we have verified that the sum of the electric-type tidal invariants, $\Delta\lambda^E_1$, $\Delta\lambda^E_2$ and $\Delta\lambda^E_3$, is zero for any pN order. We also find that the contributions to this sum from the retarded field, prior to regularization, vanishes mode-by-mode for each $\ell$.

Because of the great length of the final expressions, we place the analytical coefficients in Appendix~\ref{appB} and the numerical coefficients in the supplementary text files.

\section{Discussion}
In this article, we have developed and implemented tools for computing invariant quantities in a ``no-string" completed radiation gauge. At the level of formalism, we have defined a Detweiler-Whiting regular field in the no-string gauge, and we have shown that the invariants constructed from this field take the same value as they would if constructed from the DW field in any gauge smoothly related to Lorenz. At the level of practical calculation, we have described a simple mode-sum regularization scheme involving averaging from two sides of the particle. In the cases of the spin-precession and tidal invariants, the scheme is encapsulated in Eqs.~\eqref{psi-modesum-final}--\eqref{lambda-modesum-final}. The validity of this scheme will be rigorously established in a companion paper. We have also identified continuity conditions that must be imposed on the completion part of the metric perturbation in order to obtain the correct values for the invariants, and we expect that an extension of these conditions will be vital to calculations in a Kerr background.

By applying our regularization scheme, we obtained our main results: very high order pN expansions of the linear-in-mass-ratio corrections to the spin-precession and tidal invariants for a particle in circular orbit about a Schwarzschild black hole. These expansions were extracted from numerical results at very large orbital radii, from $10^{18}M \textrm{ to } 9\times10^{33} M$. Using the analytical solution to the Teukolsky equation found in Ref.~\cite{MST}, we were able to obtain these numerical results with accuracy greater than 1 part in $10^{500}$. This data was then used to numerically fit and extract the analytical and numerical coefficients in the pN expansion of the invariants. 

To increase the accuracy of these numerically calculated quantities (at such large radii), one would have to drastically increase the maximum computed $\ell$ mode.
This difficulty arises because of the large-$\ell$ behavior of the singular field, which makes the mode sum converge slowly unless one subtracts both the positive-power-of-$L$ terms and a large number of tail terms. The situation here contrasts with calculations of fluxes, in which the singular field plays no role. To compute fluxes with an accuracy of 1 part in $10^{600}$ in Ref.~\cite{SKerr}, going to $\ell_\textrm{max}=40$ was sufficient, whereas to compute conservative invariants here, we went to $\ell_\textrm{max}=200$. 

Instead of hunting greater accuracy, the next step will be to calculate the conservative invariants, along with the pN expansion of the Detweiler redshift, in Kerr spacetime. This should provide a substantial improvement to current EOB models. A computation of spin precession for a particle in circular orbit about a Kerr black hole will provide the linear-in-mass-ratio, spin-spin part of the effective Hamiltonian and improve the EOB model for tidally interacting spinning binary systems.



\begin{acknowledgments}
We thank Barry Wardell, Chris Kavanagh, and Adrian Ottewill for providing a detailed comparison of their pN expansions with our own. We also thank Barry Wardell and Leor Barack for confirming (and explaining) several properties of the harmonic decomposition of the Lorenz-gauge singular field, Maarten van de Meent for assisting us in identifying conical singularities, and Leor Barack, Nathan K. Johnson-McDaniel, and Sam Dolan for careful reading of the manuscript. This work was supported by the European Research Council under the European Union's Seventh Framework Programme (FP7/2007-2013)/ERC grant agreement no. 304978. 
\end{acknowledgments}

\appendix
\section{Metric perturbation corresponding to the change in mass of the spacetime}
\label{appA}

The $\ell=0$ perturbation we used in Sec.~\ref{completion} was suitable for our purposes, but by transforming to ingoing Eddington-Finkelstein (EF) coordinates, one can see that it is irregular at the future horizon. In addition, it is discontinuous across $r=r_0$. Such gauge singularities do not affect the calculation of invariants, but it is worthwhile to construct the perturbation in a gauge that is manifestly regular. Since, it is easy to check the regularity of the metric at the future event horizon in the ingoing EF coordinates ($v,r,\theta,\phi$), we start with the Schwarzschild metric in the those coordinates:
\begin{align} \label{Schwmetric}
ds^2 = -\left(1-\frac{2M}{r}\right)dv^2 + 2dvdr +r^2d\theta^2 + r^2\sin^2\theta d\phi^2.
\end{align}

We know from the solution in Sec.~\ref{completion} that across $r=r_0$, there is an invariant shift of the spacetime's mass by an amount $\mu \hat E$, and everything other than the mass in the perturbation is gauge. (The jump in mass can also be discovered from the invariant definitions of mass in Ref.~\cite{DB-Schw}.)  So we begin with a straightforward variation of the mass in Eq.~\eqref{Schwmetric},
\begin{align}\label{dMEF}
h^\textrm{EF}_{vv} = \frac{2\mu E}{r} \theta(r-r_0),
\end{align}
with the rest of the components being zero. Note that the linearized Einstein equation only tells us about the \emph{jump} in mass across the particle's orbit; it does not tell us anything about the total mass, which can be freely altered by adding a trivial perturbation of $M$ everywhere in spacetime. Here we have chosen to set the shift in  mass to be zero at $r<r_0$.

We now perform a gauge transformation, $h^\textrm{EF}_{ab} \rightarrow h_{ab} = h^\textrm{EF}_{ab} + \nabla_a\xi_b + \nabla_b\xi_a$, so that $h_{ab}$ is continuous across $r=r_0$, regular at the horizon, and asymptotically flat at spatial infinity. Our strategy is similar to that of Ref.~\cite{lpthesis}. We choose a spherically symmetric gauge vector (which respects the symmetry of the problem), 
\begin{align} \label{gv}
\xi^a = (P(v,r), Q(v,r),0,0).
\end{align}
The term, $\Xi_{ab} :=\nabla_a \xi_b + \nabla_b \xi_a$, is
\begin{widetext}
\[  \Xi_{ab} = \left( \begin{array}{cccc}
\left(2\dot{Q}-\frac{2M}{r^2}Q-2f\dot{P}\right)				&			\left(\dot{P}+Q^\prime-f P^\prime\right) 		& 			0 				& 				0			\\
\left(\dot{P}+Q^\prime-f P^\prime\right) 					&			2P^\prime 							& 			0 				& 				0			\\
0 												&				0 								& 			2rQ 			& 				0 			\\
0 												&				0 								& 			 0 				& 				2rQ\sin^2\theta 
\end{array} \right),\]
\end{widetext}
where a dot and a prime denote derivatives with respect to $v$ and $r$, respectively, and $f=(1-2M/r)$. To preserve the form of Eq (\ref{Schwmetric}), we set $Q=Q(r)$ and $P=P(v)$. Using this, we now have,
\begin{widetext}
\[ \Xi_{ab} = \left( \begin{array}{cccc}
\left(-\frac{2M}{r^2}Q-2f\dot{P}\right)			&			\left(\dot{P}+Q^\prime\right)					& 			0 				& 				0			\\
\left(\dot{P}+Q^\prime\right) 					&				0 								& 			0 				& 				0			\\
0 									&				0 								& 			2rQ 			& 				0 			\\
0 									&				0 								& 			 0 				& 				2rQ\sin^2\theta 
\end{array} \right).\]
\end{widetext}
To make the metric perturbation independent of the $v$ coordinate, we have $P=\alpha\,v$ where $\alpha$ is a constant which needs to be determined from the three conditions mentioned earlier. 
We now split our gauge vector in half, $\xi = \xi_< \theta(r_0-r) + \xi_> \theta(r-r_0)$. Since we want a regular perturbation at the horizon, we choose Q on the inside to have the form $(r-2M)^i$, where $i$ is a positive integer which will be fixed later, and to guarantee flatness at spatial infinity we choose Q to have the form $1/(r-2M)^j$, with $j$ a positive integer which will be fixed later. Without loss of generality, we write
\begin{align}
\xi_< &= \left( \alpha_< v, \beta_< \left(\frac{r-2M}{r_0-2M}\right)^i,0,0 \right), \nonumber \\
\xi_> &= \left( \alpha_> v, \beta_>\left(\frac{r_0-2M}{r-2M}\right)^j,0,0 \right),
\end{align}
in which $\alpha_{</>}$ and $\beta_{</>}$ need to be calculated. Imposing continuity of $h_{\theta\theta}$ or $h_{\phi\phi}$ at $r=r_0$ tells us that
\begin{align}
\beta_< = \beta_> = \beta.
\end{align}
Imposing the asymptotic flatness condition gives us $\alpha_>=0$. Imposing continuity of $h_{vv}$ fixes $\alpha_<$,
\begin{align}
\alpha_<= \frac{-\mu \hat E}{r_0 f_0},
\end{align}
and imposing continuity of $h_{vr}$ gives us
\begin{align}
\beta = \frac{\mu \hat E}{i+j}.
\end{align}

We have thus calculated the metric perturbation which is regular at the horizon, continuous across $r=r_0$ and flat at spatial infinity. We now write this metric perturbation in Schwarzschild coordinates $(t,r,\theta,\phi)$. For $r<r_0$,
\begin{align}
h_{tt}&=2\hat E\mu \left[ \frac{1-\frac{2M}{r}}{r_0-2M} - \frac{M}{r^2(i+j)} \left(\frac{r-2M}{r_0-2M}\right)^i \right], \nonumber \\
h_{tr}&= \hat E\mu \left[ \frac{1}{r_0-2M} + \frac{(i\,r-2M)}{r(i+j)(r-2M)} \left( \frac{r-2M}{r_0-2M} \right)^i \right], \nonumber \\
h_{rr}&= \frac{2\hat E\mu(i\,r-M)}{(i+j)(r-2M)^2} \left( \frac{r-2M}{r_0-2M} \right)^i, \nonumber \\
h_{\theta\theta}&= \frac{2r\hat  E \mu}{(i+j)}\left( \frac{r-2M}{r_0-2M} \right)^i, \nonumber \\
h_{\phi\phi}&=\frac{2r \hat E \mu \sin^2\theta}{(i+j)}\left( \frac{r-2M}{r_0-2M} \right)^i.
\end{align}
For this metric perturbation to be regular at the future event horizon, $i\ge2$. For $r>r_0$,
\begin{align}
h_{tt}&=2\hat E\mu \left[ \frac{1}{r} - \frac{M}{r^2(i+j)} \left(\frac{r_0-2M}{r-2M}\right)^j \right], \nonumber \\
h_{tr}&= \hat E\mu \left[ \frac{2}{r-2M} - \frac{(j\,r+2M)}{r(i+j)(r-2M)} \left( \frac{r_0-2M}{r-2M} \right)^j \right], \nonumber \\
h_{rr}&= 2\hat E\mu \left[ \frac{r}{(r-2M)^2} - \frac{(j\,r+M)}{(i+j)(r-2M)^2} \left( \frac{r_0-2M}{r-2M} \right)^j \right], \nonumber \\
h_{\theta\theta}&= \frac{2r \hat E \mu}{(i+j)}\left( \frac{r_0-2M}{r-2M} \right)^j, \nonumber \\
h_{\phi\phi}&=\frac{2r \hat E \mu \sin^2\theta}{(i+j)}\left( \frac{r_0-2M}{r-2M} \right)^j.
\end{align}
For this metric perturbation, $j\ge2$. We find that this metric perturbation is not manifestly static, having a nonzero time-independent $h_{tr}$ component. Of course, this is simply a gauge artefact, as we know from the solution in Sec.~\ref{completion} that the metric is static; in the gauge of this section, the timelike, hypersurface-orthogonal Killing vector is simply no longer $t^\mu$ but some appropriate $t^\mu+\delta t^\mu$.

The perturbation we have found is actually a large class of (physically equivalent) perturbations, corresponding to a variety of gauges labelled by $i$ and $j$. However, no two members of this class are related by the helically asymmetric gauge vectors discussed in Sec.~\ref{completion}, and in line with that discussion, we have found experimentally that the final result for any of the invariants is independent of the values of $i$ and $j$. Furthermore, again in line with Sec.~\ref{completion}, we have also verified experimentally that continuity of the components of the $\ell=0$ solution need not be imposed \emph{except} on $h_{vv}$ (or $h_{tt}$); discontinuity of other components does not affect the invariants. 

It is not difficult to see that using the spherically symmetric gauge vector given in Eq. (\ref{gv}), it is not possible to find a gauge in which the metric perturbation is manifestly continuous, regular at the horizon, asymptotically flat, and static. This agrees with the extensive analysis (focused on the Lorenz gauge) by Dolan and Barack in Ref.~\cite{DB-Schw}. Perhaps surprisingly, this appears to stem from our choice of total mass of the spacetime. There do exist $\ell=0$ solutions with all the above desired properties, but with a total mass of $M+dM+\mu\hat E$ rather than $M+\mu \hat E$. These slututions, first discovered in the Lorenz gauge by Berndtson~\cite{Berndtson:07}, correspond to having transplanted a small part of the black hole's mass into the perturbation, such that the mass of the background spacetime, $M$, differs from the black hole's physical mass, $M_{BH}=M+dM$. While one can use these solutions, one must be mindful that $M_{BH}\neq M$ in order to avoid any hiccups.



\section{Analytical results}
\label{appB}
Most terms in our pN expansions are obtained as real numbers with a finite numerical precision. However, many terms we find in exact, analytical form. In this appendix we present those analytical coefficients. The numerical coefficients are presented in the accompanying text files. 

We present the analytical coefficients of $\delta \psi$ for the following pN orders: (i) all the coefficients that were already known, which go up to $y^{17/2}$~\cite{BD-spin} (as later corrected in Ref.~\cite{BD-spin-new}), (ii) the highest power of $\log(y)$ for each pN order from $y^9$ to $y^{37/2}$, and (iii) the second-highest power of $\log(y)$ for each integer-pN-order from $y^9$ to $y^{15}$. 
\begin{widetext}
\begin{align} \allowdisplaybreaks
&\Delta\psi =
y^2
-3y^3
-\frac{15}{2}y^4
+ \left( \frac{-6277}{30} - \frac{496 \log(2)}{15} - 16\gamma + \frac{20471 \pi^2}{1024} -8 \log(y) \right) y^5
+ \left( -\frac{87055}{28} + \frac{653629 \pi ^2}{2048} -\frac{52 \gamma }{5} \right. \nonumber \\ & \left. + \frac{3772 \log (2)}{105} -\frac{729 \log (3)}{14} -\frac{26 \log (y)}{5} \right) y^6
-\frac{26536 \pi }{1575} y^{13/2}
+ \left( -\frac{149628163}{18900} + \frac{7628 \gamma }{21} + \frac{297761947 \pi ^2}{393216} -\frac{1407987 \pi ^4}{524288} \right. \nonumber \\ & \left. + \frac{4556 \log (2)}{21} + \frac{12879 \log (3)}{35} +\frac{3814 \log (y)}{21} \right) y^7
-\frac{113411 \pi }{22050} y^{15/2}
+\left( \frac{403109158099}{9922500} -\frac{74909462 \gamma }{70875} + \frac{3424 \gamma ^2}{25} \right. \nonumber \\ & \left. + \frac{164673979457 \pi ^2}{353894400} -\frac{160934764317 \pi ^4}{335544320} + \frac{340681718 \log (2)}{1819125}  + \frac{869696 \gamma  \log (2)}{1575} + \frac{58208 \log ^2(2)}{105} -\frac{199989 \log (3)}{352} \right. \nonumber \displaybreak\\ & \left. - \frac{9765625 \log (5)}{28512} -\frac{1344 \zeta (3)}{5} -\frac{37454731 \log (y)}{70875} + \frac{434848 \log (2) \log (y)}{1575} + \frac{3424}{25} \gamma  \log (y) + \frac{856 \log ^2(y)}{25} \right) y^8 \nonumber \\   &
+ \frac{1179591206 \pi }{3274425} y^{17/2}
- \left( \frac{2227389947 \log (y)}{606375} -\frac{3376}{15} \gamma  \log (y) + \frac{1538864 \log (2) \log (y)}{11025} -\frac{28431}{49} \log (3) \log (y) \right. \nonumber \\ & \left. -\frac{844}{15} \log ^2(y) \right) y^9
+\frac{23264368 \pi  \log (y)}{165375} y^{19/2}
-\left( -\frac{5724079403437 \log (y)}{496621125} + \frac{35570296 \gamma  \log (y)}{11025} + \frac{239211992 \log (2) \log (y)}{654885} \right. \nonumber \\ & \left. + \frac{54832464 \log (3) \log (y)}{13475} + \frac{8892574 \log ^2(y)}{11025} \right) y^{10}
+ \frac{23162039 \pi  \log (y)}{128625} y^{21/2}
+\left( \frac{1059800138707 \log ^2(y)}{756392175} -\frac{6228256 \gamma  \log ^2(y)}{11025} \right. \nonumber \\ & \left. - \frac{26744864 \log (2) \log ^2(y)}{23625} -\frac{3114128 \log ^3(y)}{33075} \right) y^{11}
-\frac{7216298231038 \pi  \log (y)}{2269176525}  y^{23/2}
+ \left( \frac{30579194748876386 \log ^2(y)}{1491353238375} \right. \nonumber  \\ & \left. -\frac{690015272 \gamma  \log ^2(y)}{385875} -\frac{14320872 \log (2) \log ^2(y)}{42875} -\frac{1108809}{343} \log (3) \log ^2(y) -\frac{345007636 \log ^3(y)}{1157625} \right) y^{12}  \nonumber \\ &
-\frac{1430850224 \pi  \log ^2(y)}{2480625} y^{25/2}
+ \left( -\frac{58168822370617659002 \log ^2(y)}{2214659558986875} + \frac{48902305452872 \gamma  \log ^2(y)}{3781960875}  \right. \nonumber \\ & \left. -\frac{18295600819864 \log (2) \log ^2(y)}{1620840375} + \frac{116958256614 \log (3) \log ^2(y)}{5187875} + \frac{24451152726436 \log ^3(y)}{11345882625} \right) y^{13} \nonumber  \\ &
-\frac{187767567979 \pi  \log ^2(y)}{121550625} y^{27/2}
+ \left( -\frac{1857799285622072434 \log ^3(y)}{33219893384803125}  + \frac{26735338432 \gamma  \log ^3(y)}{17364375} \right. \nonumber \\ & \left. + \frac{160490433344 \log (2) \log ^3(y)}{52093125} + \frac{3341917304 \log ^4(y)}{17364375} \right) y^{14}
+ \frac{504471961630030612 \pi  \log ^2(y)}{39313483295625} y^{29/2} \nonumber \\ &
+ \left( -\frac{10005212135393611353494 \log ^3(y)}{143952871334146875} + \frac{161045632 \gamma  \log ^3(y)}{19845} + \frac{1537620535168 \log (2) \log ^3(y)}{364651875} \right. \nonumber \\ & \left. + \frac{28829034 \log (3) \log ^3(y)}{2401} + \frac{20130704 \log ^4(y)}{19845} \right)  y^{15}
+ \frac{8586238183904 \pi  \log ^3(y)}{5469778125} y^{31/2}
-\frac{131271672785838026 \log ^4(y)}{39313483295625} y^{16}  \nonumber \\ &
+ \frac{276172585240238 \pi  \log ^3(y)}{38288446875} y^{33/2}
-\frac{27264557008 \log ^5(y)}{86821875}  y^{17}
-\frac{2256322610392198793612 \pi  \log ^3(y)}{81732731771604375}  y^{35/2}  \nonumber \\ &
-\frac{23312186833892 \log ^5(y)}{9116296875}  y^{18}
-\frac{1838128196186192 \pi  \log ^4(y)}{574326703125}  y^{37/2}
\end{align}
\end{widetext}

For $\Delta\lambda_1^\textrm{E}$, we present (i) all the pN coefficients (known from \cite{BD-tidal}) up to $y^{21/2}$, (ii) the highest power of $\log(y)$ for each pN order from $y^{11}$ to $y^{43/2}$, and (iii) the second-highest power of $\log(y)$ for each pN order from $y^{11}$ to $y^{17}$.  
\begin{widetext}
\begin{align} \allowdisplaybreaks
&\Delta\lambda_1^\textrm{E} =
2y^3
+2y^4
-\frac{19}{4}y^5
+ \left( \frac{227}{3} -\frac{593 \pi ^2}{256} \right) y^6
+ \left( -\frac{71779}{4800} -\frac{719 \pi ^2}{256} + \frac{1536 \log (2)}{5} + \frac{768 \gamma }{5} + \frac{384 \log (y)}{5} \right)y^7 \nonumber  \\ &
+ \left( \frac{35629703}{100800} -\frac{1008787 \pi ^2}{24576}  -\frac{5248 \log (2)}{7}  -\frac{17152 \gamma }{105} + \frac{2916 \log (3)}{7} -\frac{8576 \log (y)}{105} \right) y^8
+ \frac{27392 \pi }{175} y^{17/2}
+ \left( -\frac{5435624 \gamma }{2835} \right. \nonumber \\ & \left. + \frac{4692901483 \pi ^2}{7077888} + \frac{877432 \log (2)}{2835} -\frac{20898 \log (3)}{7}  + \frac{2193373 \pi ^4}{1048576} -\frac{6746904013}{7257600}  -\frac{2717812 \log (y)}{2835} \right) y^9
-\frac{254116 \pi }{1225} y^{19/2}  \nonumber \\ & 
+ \left( \frac{58241403128 \gamma }{5457375} -\frac{876544}{175} \gamma  \log (2) + \frac{113134518813241 \pi ^2}{19818086400} + \frac{6396680456 \log (2)}{5457375} + \frac{6028101 \log (3)}{1120} -\frac{6653357405 \pi ^4}{67108864}  \right. \nonumber \\ & \left. + \frac{12288 \zeta (3)}{5} + \frac{9765625 \log (5)}{3168} -\frac{876544}{175} \log ^2(2) -\frac{219136 \gamma ^2}{175} -\frac{1964481413350639}{48898080000} + \frac{29120701564 \log (y)}{5457375} \right. \nonumber \\ & \left. -\frac{219136}{175} \gamma  \log (y) -\frac{438272}{175} \log (2) \log (y) -\frac{54784}{175} \log ^2(y) \right) y^{10}
-\frac{5977039346 \pi }{3274425} y^{21/2}
+ \left( \frac{170466930577 \log (y)}{9459450} \right. \nonumber \displaybreak \\ & \left. + \frac{39488}{315} \gamma  \log (y) + \frac{18028096 \log (2) \log (y)}{3675} -\frac{227448}{49} \log (3) \log (y)  + \frac{9872 \log ^2(y)}{315} \right) y^{11}
+ \left( \frac{6426556598284309 \pi }{524431908000} \right. \nonumber \\ & \left. -\frac{46895104 \gamma  \pi }{18375} + \frac{438272 \pi ^3}{525} -\frac{93790208 \pi  \log (2)}{18375} -\frac{23447552 \pi  \log (y)}{18375} \right) y^{23/2}
+ \left( -\frac{53463358651678127 \log (y)}{884978844750} \right. \nonumber \\ & \left. + \frac{169954100048 \gamma  \log (y)}{9823275} -\frac{30332610032 \log (2) \log (y)}{1964655} + \frac{1630044}{49} \log (3) \log (y) + \frac{42488525012 \log ^2(y)}{9823275} \right) y^{12} \nonumber \\ &
+ \left( \frac{425532494729325719 \pi }{14913532383750}  + \frac{427856752 \gamma  \pi }{385875} + \frac{16089104 \pi ^3}{11025} + \frac{12561842768 \pi  \log (2)}{1157625} -\frac{2956824}{343} \pi  \log (3) \right. \nonumber \\ & \left. + \frac{213928376 \pi  \log (y)}{385875} \right) y^{25/2}
+ \left( -\frac{815629077100102 \log ^2(y)}{49165491375} + \frac{93790208 \gamma  \log ^2(y)}{18375} + \frac{187580416 \log (2) \log ^2(y)}{18375} \right. \nonumber \\ & \left. + \frac{46895104 \log ^3(y)}{55125} \right) y^{13}
+ \left( -\frac{425577146707570254105031 \pi }{3212732800236960000} + \frac{928193260264 \gamma  \pi }{28014525} -\frac{30834184 \pi ^3}{2695} \right. \nonumber \\ & \left. -\frac{80246336987576 \pi  \log (2)}{3094331625} + \frac{21190572}{343} \pi  \log (3) + \frac{464096630132 \pi  \log (y)}{28014525} \right) y^{27/2}
+ \left( -\frac{235234718516611549 \log ^2(y)}{2130504626250} \right. \nonumber \\ & \left. + \frac{1056674176 \gamma  \log ^2(y)}{165375} -\frac{1684425088 \log (2) \log ^2(y)}{128625} + \frac{8870472}{343} \log (3) \log ^2(y) + \frac{528337088 \log ^3(y)}{496125} \right) y^{14}  \nonumber  \\ &
+ \left( -\frac{4846441673897531348153 \pi  \log (y)}{59057588239650000} + \frac{40142209024 \gamma  \pi  \log (y)}{1929375} -\frac{375160832 \pi ^3 \log (y)}{55125} + \frac{80284418048 \pi  \log (2) \log (y)}{1929375} \right. \nonumber  \\ & \left. + \frac{10035552256 \pi  \log ^2(y)}{1929375} \right) y^{29/2}
+ \left( \frac{25596917755268555111989 \log ^2(y)}{387249042885705000} -\frac{326830998824752 \gamma  \log ^2(y)}{4862521125}  \right. \nonumber  \\ & \left. + \frac{5873797739780528 \log (2) \log ^2(y)}{34037647875} -\frac{63571716}{343} \log (3) \log ^2(y) -\frac{163415499412376 \log ^3(y)}{14587563375} \right) y^{15} \nonumber  \\ &
+ \left( -\frac{41920257709287378783847 \pi  \log (y)}{111963344371003125} + \frac{674951603104 \gamma  \pi  \log (y)}{40516875} -\frac{9936857824 \pi ^3 \log (y)}{385875} \right. \nonumber \\ & \left. -\frac{217981327456 \pi  \log (2) \log (y)}{3472875} + \frac{230632272 \pi  \log (3) \log (y)}{2401} + \frac{168737900776 \pi  \log ^2(y)}{40516875} \right) y^{31/2} \nonumber  \\ &
+ \left( \frac{681852850098374173148 \log ^3(y)}{33219893384803125} -\frac{80284418048 \gamma  \log ^3(y)}{5788125} -\frac{160568836096 \log (2) \log ^3(y)}{5788125} \right. \nonumber \\ & \left. -\frac{10035552256 \log ^4(y)}{5788125} \right) y^{16}
+ \left( \frac{57774176073258410829735523 \pi  \log
   (y)}{124777599295357008000} -\frac{1472383886624419184 \gamma  \pi  \log (y)}{5616211899375} \right. \nonumber \\ & \left. + \frac{247867707836464 \pi ^3 \log (y)}{4862521125} + \frac{71392412350010824144 \pi  \log (2) \log (y)}{117940449886875} -\frac{1652864616 \pi  \log (3) \log (y)}{2401} \right. \nonumber \\ & \left. -\frac{368095971656104796 \pi  \log ^2(y)}{5616211899375} \right) y^{33/2}
+ \left( \frac{1170236981712855093944183 \log ^3(y)}{3023010298017084375} -\frac{3144434215808 \gamma  \log ^3(y)}{72930375} \right. \nonumber \\ & \left. + \frac{1195410487168 \log (2) \log ^3(y)}{121550625} -\frac{230632272 \log (3) \log ^3(y)}{2401} -\frac{393054276976 \log ^4(y)}{72930375} \right) y^{17}
-\frac{8590432731136 \pi  \log ^3(y)}{607753125} y^{35/2} \nonumber  \\ &
+ \frac{4868681792431448132  \log ^4(y)}{353821349660625} y^{18}
-\frac{49975432673168 \pi   \log ^3(y)}{1418090625} y^{37/2}
+\frac{8590432731136  \log ^5(y)}{3038765625} y^{19}  \nonumber \\ &
+ \frac{9497909709044998367528 \pi   \log^3(y)}{81732731771604375} y^{39/2}
+ \frac{436689749341952  \log ^5(y)}{27348890625} y^{20}
+ \frac{1838352604463104 \pi   \log ^4(y)}{63814078125} y^{41/2} \nonumber \\ &
+ \frac{119724959062139341591288 \log^5(y)}{6129954882870328125} y^{21}
+ \frac{185606588227076264 \pi  \log ^4(y)}{1340095640625} y^{43/2}
\end{align}
\end{widetext}

For $\Delta\lambda_2^\textrm{E}$, we present (i) all the pN coefficients (known from \cite{BD-tidal}) up to $y^{19/2}$, (ii) the highest power of $\log(y)$ for each pN order from $y^{10}$ to $y^{43/2}$, and (iii) the second-highest power of $\log(y)$ for each pN order from $y^{10}$ to $y^{17}$.
\begin{widetext}
\begin{align} \allowdisplaybreaks
&\Delta\lambda_2^\textrm{E} =
-y^3
-\frac32 y^4
-\frac{23}{8}y^5
+ \left( -\frac{2593}{48} + \frac{1249 \pi ^2}{1024} \right) y^6
+ \left( -\frac{362051}{3200}  + \frac{1737 \pi ^2}{1024} -\frac{256 \gamma }{5} -\frac{512 \log (2)}{5} -\frac{128 \log (y)}{5} \right) y^7 \nonumber \\ &
+   \left( \frac{917879}{1280} -\frac{7637151 \pi ^2}{65536} + \frac{16592 \log (2)}{105} + \frac{176 \gamma }{7} -\frac{729 \log (3)}{7} + \frac{88 \log (y)}{7} \right) y^8
-\frac{27392 \pi }{525} y^{17/2}
+ \left( \frac{1193824 \gamma }{2835} \right. \nonumber \\ & \left. -\frac{24327985735 \pi ^2}{14155776} + \frac{2368 \log (2)}{405} + \frac{1215 \log (3)}{2} + \frac{29225393 \pi ^4}{2097152} + \frac{35725395527}{2903040} + \frac{596912 \log (y)}{2835} \right) y^9
+ \frac{58087 \pi }{1575} y^{19/2} \nonumber \\ &
+ \left( -\frac{3286454596 \log (y)}{1819125} + \frac{219136}{525} \gamma  \log (y) + \frac{438272}{525} \log (2) \log (y) + \frac{54784 \log ^2(y)}{525} \right) y^{10}
+ \left( \frac{2672297839 \pi }{6548850}  \right) y^{21/2} \nonumber \\ &
+ \left( -\frac{4919133341971 \log (y)}{993242250}  + \frac{931408 \gamma  \log (y)}{11025} -\frac{1573072 \log (2) \log (y)}{1575} + \frac{56862}{49} \log (3) \log (y) + \frac{232852 \log ^2(y)}{11025} \right) y^{11}  \nonumber  \\ &
+ \left( -\frac{1425524472919397 \pi }{349621272000} + \frac{46895104 \gamma  \pi }{55125} -\frac{438272 \pi ^3}{1575} + \frac{93790208 \pi  \log (2)}{55125} + \frac{23447552 \pi  \log (y)}{55125} \right) y^{23/2} \nonumber \\ &
+ \left( \frac{336663160031017 \log (y)}{25744839120} -\frac{3437020472 \gamma  \log (y)}{893025} + \frac{3397300136 \log (2) \log (y)}{1403325} -\frac{47385}{7} \log (3) \log (y) \right. \nonumber  \\ & \left. -\frac{859255118 \log ^2(y)}{893025} \right) y^{12}
+ \left( -\frac{2608988253558091529 \pi }{318155357520000} -\frac{1940124 \gamma  \pi }{42875} -\frac{14576396 \pi ^3}{33075} -\frac{2608166516 \pi  \log (2)}{1157625} \right. \nonumber  \\ & \left. + \frac{739206}{343} \pi  \log (3) -\frac{970062 \pi  \log (y)}{42875} \right) y^{25/2}
+ \left( \frac{72036007252477 \log ^2(y)}{10925664750} -\frac{93790208 \gamma  \log ^2(y)}{55125} -\frac{187580416 \log (2) \log ^2(y)}{55125} \right. \nonumber \\ & \left. -\frac{46895104 \log ^3(y)}{165375} \right) y^{13}
+ \left( \frac{336629234198810065383583 \pi }{11565838080853056000} -\frac{84454061934086 \gamma  \pi }{11345882625} + \frac{25421917126 \pi ^3}{9823275} \right. \nonumber \\ & \left. + \frac{731692656242 \pi  \log (2)}{194500845} -\frac{616005}{49} \pi  \log (3) -\frac{42227030967043 \pi  \log (y)}{11345882625} \right) y^{27/2}
+ \left( \frac{2680194891812081201 \log ^2(y)}{89481194302500} \right. \nonumber  \\ & \left.-\frac{2381473984 \gamma  \log ^2(y)}{1157625} + \frac{910030784 \log (2) \log ^2(y)}{385875} -\frac{2217618}{343} \log (3) \log ^2(y) -\frac{1190736992 \log ^3(y)}{3472875} \right) y^{14}  \nonumber \\ &
+ \left( \frac{3627715358766651040027 \pi  \log (y)}{118115176479300000} -\frac{40142209024 \gamma  \pi  \log (y)}{5788125} + \frac{375160832 \pi ^3 \log (y)}{165375} -\frac{80284418048 \pi  \log (2) \log (y)}{5788125} \right. \nonumber  \\ & \left. -\frac{10035552256 \pi  \log ^2(y)}{5788125} \right) y^{29/2}
+ \left( -\frac{230100400220601639975419 \log ^2(y)}{10842973200799740000} + \frac{104146032969424 \gamma  \log ^2(y)}{6807529575} \right. \nonumber \\ & \left. -\frac{1070223775519216 \log (2) \log ^2(y)}{34037647875} + \frac{1848015}{49} \log (3) \log ^2(y) + \frac{52073016484712 \log ^3(y)}{20422588725} \right) y^{15} \nonumber \\ &
+ \left( \frac{959470947789066888806999 \pi  \log (y)}{9212983765385400000} -\frac{104717532296 \gamma  \pi  \log (y)}{17364375} + \frac{1165602776 \pi ^3 \log (y)}{165375} + \frac{161636407304 \pi  \log (2) \log (y)}{13505625} \right. \nonumber \\ & \left. -\frac{57658068 \pi  \log (3) \log (y)}{2401} -\frac{26179383074 \pi  \log ^2(y)}{17364375} \right) y^{31/2}
+ \left( -\frac{151808990580596480372 \log ^3(y)}{11073297794934375} + \frac{80284418048 \gamma  \log ^3(y)}{17364375} \right. \nonumber \\ & \left. + \frac{160568836096 \log (2) \log ^3(y)}{17364375} + \frac{10035552256 \log ^4(y)}{17364375} \right) y^{16}
+ \left( -\frac{3854184236032319074099692403 \pi  \log(y)}{30372818923129082400000} \right. \nonumber \\ & \left. + \frac{87324664477990228 \gamma  \pi  \log (y)}{1456054936875} -\frac{49560820650364 \pi ^3 \log (y)}{3781960875} -\frac{165995504947200028 \pi  \log (2) \log (y)}{1531694154375} \right. \nonumber \\ & \left. + \frac{48048390}{343} \pi  \log (3) \log (y) + \frac{21831166119497557 \pi  \log ^2(y)}{1456054936875} \right) y^{33/2}
+ \left( -\frac{334487077678301085190949 \log ^3(y)}{3023010298017084375} \right. \nonumber \\ & \left. + \frac{487016857696 \gamma  \log ^3(y)}{40516875} + \frac{7650935648 \log (2) \log ^3(y)}{364651875} + \frac{57658068 \log (3) \log ^3(y)}{2401} + \frac{60877107212 \log ^4(y)}{40516875} \right) y^{17}  \nonumber \\ &
+ \frac{8590432731136 \pi   \log ^3(y)}{1823259375}y^{35/2}
-\frac{1231646038737086398 \log ^4(y)}{353821349660625}y^{18} 
+ \frac{55101325697524 \pi  \log ^3(y)}{5469778125}y^{37/2}  \nonumber \\ &
-\frac{8590432731136  \log ^5(y)}{9116296875} y^{19}
-\frac{11871449505595480790806 \pi   \log^3(y)}{408663658858021875}y^{39/2}
-\frac{812555208074128  \log ^5(y)}{191442234375} y^{20} \nonumber \\ &
-\frac{1838352604463104 \pi   \log ^4(y)}{191442234375}y^{41/2}
-\frac{3091229654826478747312  \log^5(y)}{1225990976574065625}y^{21}
-\frac{49849519190246954 \pi   \log ^4(y)}{1340095640625} y^{43/2}
\end{align}
\end{widetext}

For $\Delta\lambda_3^\textrm{E}$, we present (i) all the pN coefficients (known from \cite{BD-tidal}) up to $y^{19/2}$, (ii) the highest power of $\log(y)$ for each pN order from $y^{10}$ to $y^{43/2}$, and (iii) the second-highest power of $\log(y)$ for each pN order from $y^{10}$ to $y^{17}$.
\begin{widetext}
\begin{align} \allowdisplaybreaks
&\Delta\lambda_3^\textrm{E} =
-y^3 
- \frac12 y^4 
+ \frac{61}{8} y^5
+ \left( \frac{1123 \pi ^2}{1024}-\frac{1039}{48} \right) y^6
+ \left( -\frac{256 \log (y)}{5}+\frac{1139 \pi ^2}{1024}-\frac{512 \gamma}{5}+\frac{1229711}{9600}-\frac{1024 \log (2)}{5} \right) y^7 \nonumber \\ &
+ \left( \frac{7256 \log (y)}{105}+\frac{30981749 \pi ^2}{196608}+\frac{14512\gamma }{105}-\frac{431650697}{403200}-\frac{2187 \log(3)}{7}+\frac{62128 \log (2)}{105} \right) y^{8}
-\frac{54784 \pi }{525} y^{17/2} \nonumber \\ &
+ \left( \frac{424180 \log (y)}{567}-\frac{33612139 \pi^4}{2097152}+\frac{14942182769 \pi ^2}{14155776}+\frac{848360\gamma }{567}-\frac{165133169609}{14515200}+\frac{33291 \log(3)}{14}-\frac{894008 \log (2)}{2835} \right) y^{9} \nonumber  \\ &
+ \frac{376087 \pi }{2205} y^{19/2}
- \left( -\frac{109568}{525} \log ^2(y)-\frac{876544}{525} \log (2) \log(y)-\frac{438272}{525} \gamma  \log (y)+\frac{19261337776 \log(y)}{5457375} \right) y^{10} \nonumber \\ &
+ \frac{3093926951 \pi }{2182950} y^{21/2}
- \left( \frac{578372 \log ^2(y)}{11025}-\frac{170586}{49} \log (3) \log(y)+\frac{43072784 \log (2) \log (y)}{11025}+\frac{2313488 \gamma\log (y)}{11025} \right. \nonumber \\ & \left. +\frac{6489947184307 \log (y)}{496621125} \right) y^{11}
+ \left( \frac{46895104 \pi  \log (y)}{55125}-\frac{876544 \pi^3}{1575}+\frac{93790208 \gamma  \pi}{55125}-\frac{8576539777810427 \pi}{1048863816000} \right. \nonumber \\ & \left. +\frac{187580416 \pi  \log (2)}{55125} \right) y^{23/2}
- \left( \frac{33036718714 \log ^2(y)}{9823275}+\frac{1298349}{49} \log (3)\log (y)-\frac{127881949208 \log (2) \log(y)}{9823275} \right. \nonumber  \\ & \left. +\frac{132146874856 \gamma  \log(y)}{9823275}-\frac{335124500204895341 \log (y)}{7079830758000} \right) y^{12}
+ \left( -\frac{29313974 \pi  \log (y)}{55125}-\frac{4812988 \pi^3}{4725}-\frac{58627948 \gamma  \pi}{55125} \right. \nonumber \\ & \left. -\frac{2772444986000367347 \pi}{136352296080000}+\frac{2217618}{343} \pi  \log(3)-\frac{368654676 \pi  \log (2)}{42875} \right) y^{25/2}
+ \left( -\frac{93790208 \log ^3(y)}{165375} \right. \nonumber \\ & \left. -\frac{375160832 \log (2) \log^2(y)}{55125}-\frac{187580416 \gamma  \log^2(y)}{55125}+\frac{982934088927911 \log ^2(y)}{98330982750} \right) y^{13}
+ \left( -\frac{145732104236417 \pi  \log (y)}{11345882625}\right. \nonumber \\ & \left. +\frac{86968683554\pi ^3}{9823275}-\frac{291464208472834 \gamma  \pi}{11345882625}+\frac{459787882287862634382511 \pi}{4448399261866560000}-\frac{16878537}{343} \pi  \log(3)\right. \nonumber \\ & \left. +\frac{754663492020986 \pi  \log (2)}{34037647875} \right) y^{27/2}
+ \left( -\frac{30958304 \log ^3(y)}{42875}-\frac{6652854}{343} \log (3) \log^2(y)+\frac{828648896 \log (2) \log ^2(y)}{77175} \right. \nonumber \\ & \left. -\frac{185749824\gamma  \log ^2(y)}{42875}+\frac{7199663285885603857 \log^2(y)}{89481194302500} \right) y^{14}
- \left( \frac{20071104512 \pi  \log ^2(y)}{5788125}+\frac{160568836096 \pi \log (2) \log (y)}{5788125} \right. \nonumber  \\ & \left. -\frac{750321664 \pi ^3 \log(y)}{165375}+\frac{80284418048 \gamma  \pi  \log(y)}{5788125}-\frac{673907554336490184031 \pi  \log(y)}{13123908497700000} \right) y^{29/2}
+ \left( \frac{883543413463072 \log ^3(y)}{102112943625} \right. \nonumber \\ & \left. +\frac{50635611}{343}\log (3) \log ^2(y)-\frac{4803573964261312 \log (2) \log^2(y)}{34037647875}+\frac{1767086826926144 \gamma  \log^2(y)}{34037647875} \right. \nonumber \\ & \left. -\frac{486613296926917903160273 \log^2(y)}{10842973200799740000} \right) y^{15}
- \left( \frac{64591604162 \pi  \log ^2(y)}{24310125}+\frac{172974204 \pi \log (3) \log (y)}{2401} \right. \nonumber \\ & \left. -\frac{6174618795224 \pi  \log (2) \log(y)}{121550625}-\frac{4330270808 \pi ^3 \log(y)}{231525}+\frac{258366416648 \gamma  \pi  \log(y)}{24310125} \right. \nonumber \\ & \left. -\frac{17429771806026061957846879 \pi  \log(y)}{64490886357697800000} \right) y^{31/2}
- \left( -\frac{20071104512 \log ^4(y)}{17364375}-\frac{321137672192 \log (2)\log ^3(y)}{17364375} \right. \nonumber \\ & \left. -\frac{160568836096 \gamma  \log^3(y)}{17364375}+\frac{226425878356584732032 \log^3(y)}{33219893384803125} \right) y^{16}
- \left( -\frac{1987230316366299533 \pi  \log^2(y)}{39313483295625} \right. \nonumber \\ & \left. -\frac{1316525886 \pi  \log (3) \log(y)}{2401}+\frac{58610758469076421988 \pi  \log (2) \log(y)}{117940449886875}+\frac{1289026569001972 \pi ^3 \log(y)}{34037647875} \right. \nonumber \\ & \left. -\frac{7948921265465198132 \gamma  \pi  \log(y)}{39313483295625}+\frac{2480775745377381465572630931521 \pi \log (y)}{7380594998320367023200000} \right) y^{33/2}
- \left( -\frac{1417377419972 \log ^4(y)}{364651875} \right. \nonumber\displaybreak \\ & \left. -\frac{172974204 \log (3)\log ^3(y)}{2401}+\frac{3593882397152 \log (2) \log^3(y)}{364651875}-\frac{11339019359776 \gamma  \log^3(y)}{364651875} \right. \nonumber \\ & \left. +\frac{278583301344851336251078 \log^3(y)}{1007670099339028125} \right) y^{17}
+\frac{17180865462272 \pi   \log ^3(y)}{1823259375}y^{35/2}
-\frac{3637035753694361734 \log ^4(y)}{353821349660625}  y^{18} \nonumber \\ &
+ \frac{963627402292868 \pi \log ^3(y)}{38288446875} y^{37/2}
-\frac{17180865462272  \log ^5(y)}{9116296875} y^{19}
-\frac{11872699679876503682278 \pi   \log^3(y)}{136221219619340625} y^{39/2} \nonumber \\ &
-\frac{2244273037319536  \log ^5(y)}{191442234375} y^{20}
-\frac{3676705208926208 \pi  \log ^4(y)}{191442234375} y^{41/2}
-\frac{9478982798909722532248  \log ^5(y)}{557268625715484375} y^{21} \nonumber \\ &
-\frac{1292924467017422 \pi   \log ^4(y)}{12762815625} y^{43/2}
\end{align}
\end{widetext}

For $\Delta\lambda^\textrm{B}$, we present (i) all the pN coefficients (known from \cite{BD-tidal}) up to $y^{11}$, (ii) the highest power of $\log(y)$ for each pN order from $y^{23/2}$ to $y^{43/2}$, and (iii) the second-highest power of $\log(y)$ for each pN order from $y^{23/2}$ to $y^{18}$.
\begin{widetext}
\begin{align} \allowdisplaybreaks
&\Delta\lambda^\textrm{B} =
2y^{7/2}
+3y^{9/2}
+ \frac{59 }{4}y^{11/2}
-\left( \frac{41 \pi ^2}{16} -\frac{2761}{24} \right)y^{13/2}
+ \left( -\frac{112919 \pi ^2}{3072} + \frac{1808 \gamma }{15} + 240 \log (2) + \frac{1618039}{2880} \right. \nonumber \\ & \left. + \frac{904 \log (y)}{15} \right) y^{15/2}
+ \left( \frac{2756 \gamma }{105}+ \frac{3645 \log (3)}{14} + \frac{491047651}{201600} -\frac{565685 \pi ^2}{3072} -\frac{4492 \log (2)}{21} + \frac{1378 \log (y)}{105} \right) y^{17/2}
+ \frac{856 \pi }{7} y^9 \nonumber  \\ &
+ \left( -\frac{200961 \log (3)}{140} -\frac{3881396 \gamma }{2835} -\frac{1992212 \log (2)}{2835} -\frac{26691349 \pi ^4}{524288} -\frac{454873888681}{50803200} + \frac{7377893735 \pi ^2}{3538944} \right. \nonumber \\ & \left. -\frac{1940698 \log (y)}{2835} \right) y^{19/2}
-\frac{69473 \pi }{22050} y^{10}
- \left( \frac{423235951437871681}{1760330880000} + \frac{1537376 \gamma ^2}{1575} + \frac{2047552}{525} \gamma  \log (2)-\frac{28736 \zeta (3)}{15} \right. \nonumber \\ & \left. -\frac{83360241649 \gamma }{10914750}-\frac{42496203125923 \pi^2}{2477260800}-\frac{89531499967 \pi^4}{100663296}+\frac{6139232 \log ^2(2)}{1575}-\frac{82889847697\log (2)}{10914750}-\frac{1412559 \log (3)}{1232} \right. \nonumber \\ & \left. -\frac{9765625\log (5)}{7128}  + \frac{1023776}{525} \log (2) \log (y)+\frac{1537376 \gamma  \log(y)}{1575}-\frac{83360241649 \log (y)}{21829500} + \frac{384344 \log ^2(y)}{1575}\right) y^{21/2} \nonumber \\ &
-\frac{5843221973 \pi }{4365900} y^{11}
- \left( \frac{2562124 \log ^2(y)}{11025}+\frac{142155}{49} \log (3) \log(y)-\frac{2324624 \log (2) \log (y)}{2205}+\frac{10248496 \gamma \log (y)}{11025} \right. \nonumber \\ & \left. -\frac{42972211891457 \log (y)}{2648646000} \right) y^{23/2}
+ \left( -\frac{54772016 \pi  \log (y)}{55125}+\frac{1023776 \pi^3}{1575}-\frac{109544032 \gamma  \pi}{55125}+\frac{177773734055963 \pi }{20170458000} \right. \nonumber  \\ & \left. -\frac{656897824\pi  \log (2)}{165375} \right) y^{12}
- \left( -\frac{28762016819 \log ^2(y)}{9823275}-\frac{433572021 \log (3)\log (y)}{26950}+\frac{13579931188 \log (2) \log(y)}{9823275} \right. \nonumber \\ & \left. -\frac{115048067276 \gamma  \log(y)}{9823275}+\frac{993028451553179707 \log (y)}{28319323032000} \right) y^{25/2}
+ \left( -\frac{7303267 \pi  \log (y)}{11025}+\frac{1495618 \pi^3}{945}-\frac{14606534 \gamma  \pi}{11025} \right. \nonumber \\ & \left. +\frac{5157596639147063 \pi}{180360180000}-\frac{1848015}{343} \pi  \log(3)+\frac{1061371118 \pi  \log (2)}{385875} \right) y^{13}
+ \left( \frac{328448912 \log ^3(y)}{496125}+\frac{87561952 \log (2) \log^2(y)}{11025} \right. \nonumber \\ & \left. +\frac{656897824 \gamma  \log^2(y)}{165375}-\frac{1286229465862031 \log ^2(y)}{98330982750} \right) y^{27/2}
+ \left( \frac{261156926572103 \pi  \log (y)}{22691765250}-\frac{72277933843\pi ^3}{9823275} \right. \nonumber \\ & \left. +\frac{261156926572103 \gamma  \pi}{11345882625}-\frac{182312652332925353457079 \pi}{2224199630933280000}+\frac{311318752347 \pi  \log(3)}{10375750}+\frac{7448963901653 \pi  \log (2)}{34037647875} \right) y^{14}  \nonumber\displaybreak \\ &
+\left( \frac{221190188 \log ^3(y)}{165375}+\frac{5544045}{343} \log (3)\log ^2(y)-\frac{48536984 \log (2) \log^2(y)}{385875}+\frac{442380376 \gamma  \log^2(y)}{55125} \right. \nonumber \\ & \left. -\frac{4569882002335828337 \log^2(y)}{51132111030000} \right) y^{29/2}
- \left( -\frac{4684564432 \pi  \log ^2(y)}{1157625}-\frac{562069329088 \pi \log (2) \log (y)}{17364375}+\frac{175123904 \pi ^3 \log(y)}{33075} \right. \nonumber  \\ & \left. -\frac{18738257728 \gamma  \pi  \log(y)}{1157625}+\frac{374731983268181645111 \pi  \log(y)}{5905758823965000} \right) y^{15}
+ \left( -\frac{755861078825096 \log ^3(y)}{102112943625} \right. \nonumber \\ & \left. -\frac{933956257041\log (3) \log ^2(y)}{10375750}+\frac{2193478035579632 \log (2)\log ^2(y)}{34037647875}-\frac{1511722157650192 \gamma  \log^2(y)}{34037647875} \right. \nonumber \\ & \left. +\frac{38836915123602398840281 \log^2(y)}{834074861599980000} \right) y^{31/2}
- \left( -\frac{810653775143 \pi  \log ^2(y)}{121550625}-\frac{144145170 \pi \log (3) \log (y)}{2401} \right. \nonumber \\ & \left. +\frac{815203478372 \pi  \log (2) \log(y)}{121550625}+\frac{24851128532 \pi ^3 \log(y)}{1157625}-\frac{3242615100572 \gamma  \pi  \log(y)}{121550625} \right. \nonumber \\ & \left. +\frac{5225794921492869641929199 \pi  \log(y)}{16122721589424450000} \right) y^{16}
- \left( \frac{70258666136 \log ^4(y)}{52093125}+\frac{4625762368 \log (2)\log ^3(y)}{214375} \right. \nonumber \\ & \left. +\frac{562069329088 \gamma  \log^3(y)}{52093125}-\frac{1558708913165878011359 \log^3(y)}{66439786769606250} \right) y^{33/2}
- \left( \frac{3494267335644677399 \pi  \log^2(y)}{78626966591250} \right. \nonumber \\ & \left. +\frac{669844207552887 \pi  \log (3) \log(y)}{1997331875}-\frac{24682733286580834678 \pi  \log (2) \log(y)}{117940449886875}-\frac{1187368270177238 \pi ^3 \log(y)}{34037647875} \right. \nonumber  \\ & \left. +\frac{6988534671289354798 \gamma  \pi  \log(y)}{39313483295625}-\frac{1239754823171201756997341727389 \pi \log (y)}{3690297499160183511600000} \right) y^{17}
- \left( \frac{1722436873016 \log ^4(y)}{364651875} \right. \nonumber \\ & \left. +\frac{144145170 \log (3)\log ^3(y)}{2401}+\frac{1132771565248 \log (2) \log^3(y)}{72930375}+\frac{13779494984128 \gamma  \log^3(y)}{364651875} \right. \nonumber \\ & \left. -\frac{678345937778148753230629 \log^3(y)}{2198552944012425000} \right) y^{35/2}
+ \left( -\frac{247478286688 \pi  \log ^3(y)}{22509375}-\frac{240540505566272\pi  \log (2) \log ^2(y)}{1823259375} \right. \nonumber \\ & \left. +\frac{4625762368 \pi ^3 \log ^2(y)}{214375}-\frac{494956573376 \gamma  \pi  \log^2(y)}{7503125}+\frac{259176410061430480963534651 \pi  \log^2(y)}{1330124531127517125000} \right) y^{18} \nonumber  \\ &
+ \frac{3420457033850597429 \log ^4(y)}{353821349660625} y^{37/2}
-\frac{252749329384718 \pi   \log ^3(y)}{7657689375}y^{19}
+\frac{60135126391568 \log ^5(y)}{27348890625}y^{39/2}  \nonumber \\ &
+ \frac{11368575748083640757621 \pi   \log^3(y)}{136221219619340625}y^{20}
+ \frac{2330901537975412  \log ^5(y)}{191442234375}y^{41/2}
+ \frac{857912842634576 \pi   \log ^4(y)}{38288446875}y^{21} \nonumber \\ &
+ \frac{46456527006194882616964  \log^5(y)}{6129954882870328125} y^{43/2}
\end{align}
\end{widetext}

\bibliography{sf6}

\begin{figure*} [htb]
\centering
\includegraphics[width=0.495\textwidth]{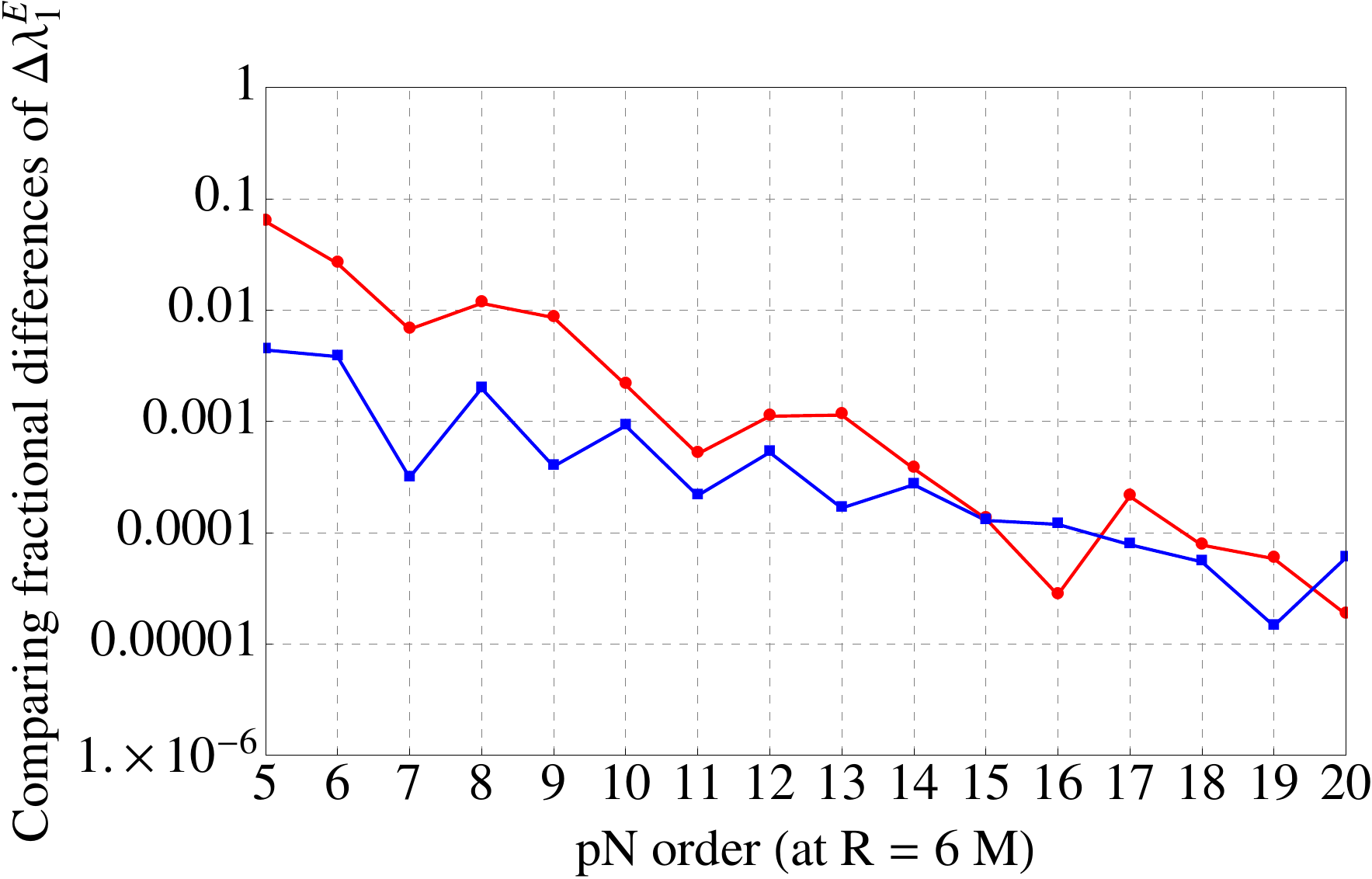}
\includegraphics[width=0.495\textwidth]{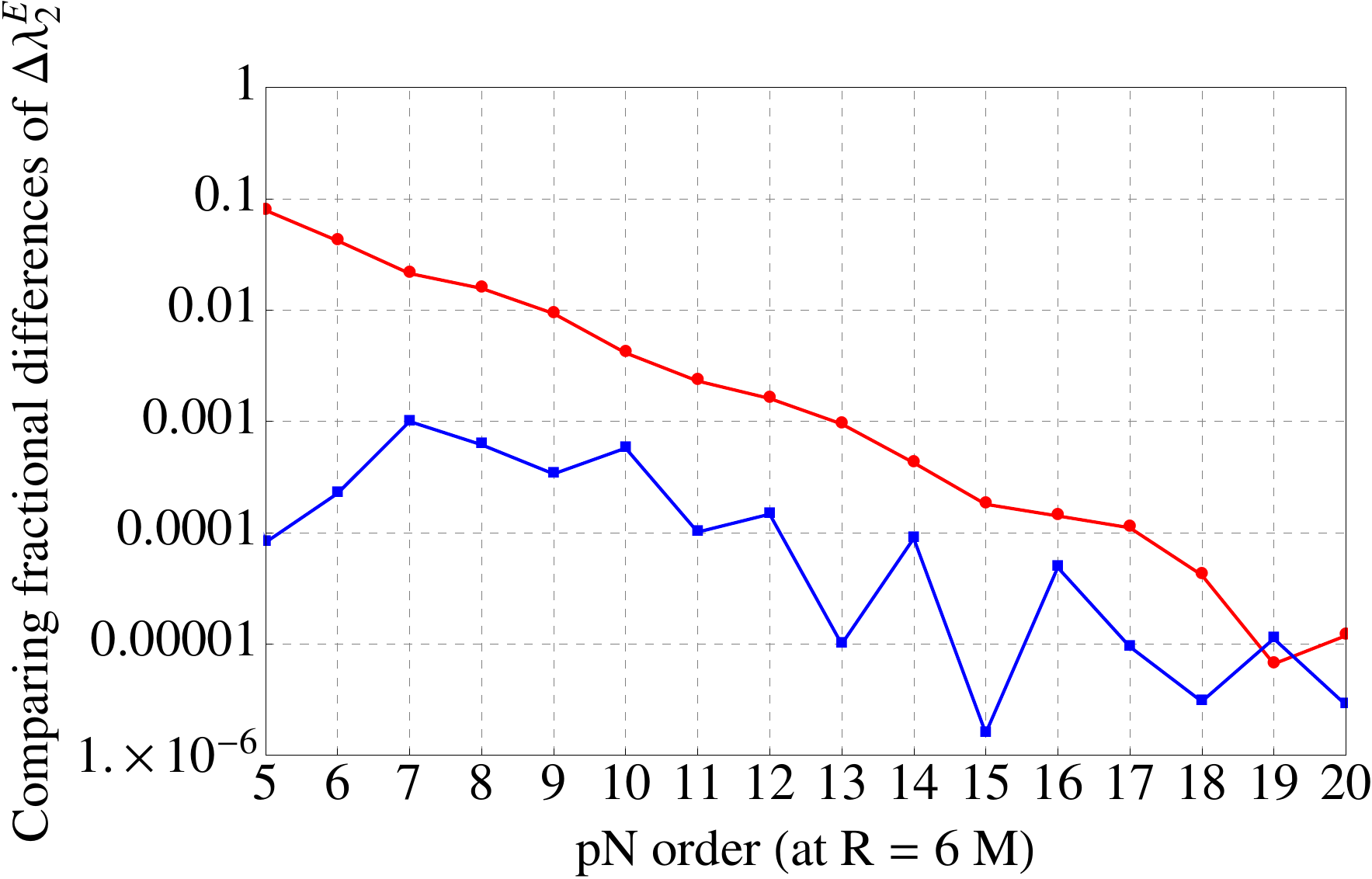}
\includegraphics[width=0.495\textwidth]{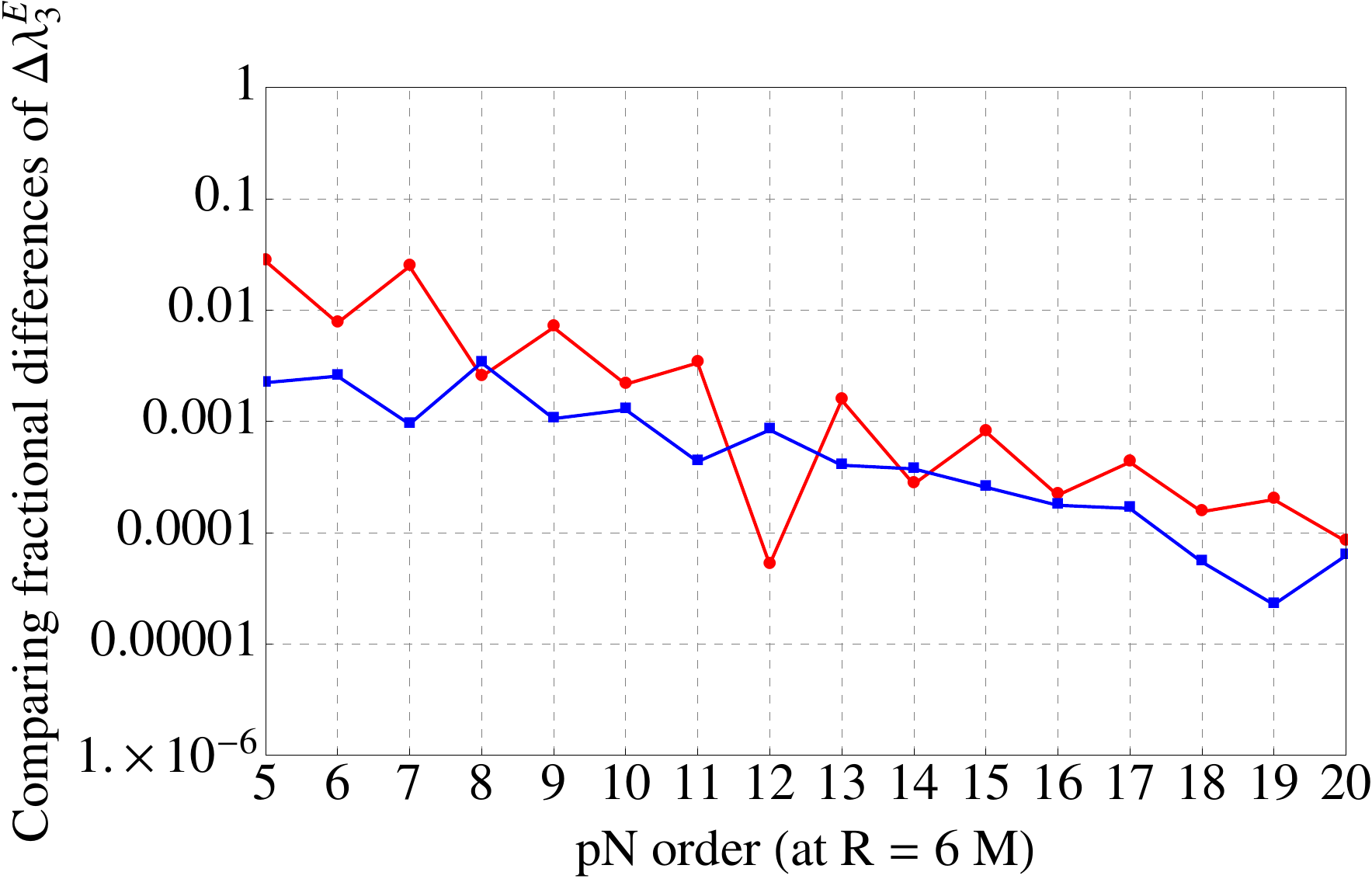}
\caption{Fractional difference between the  numerically calculated value and Taylor-summed pN series (represented by red line), and the fractional difference between the numerically calculated value and exponentially-resummed pN series (represented by blue line), for different pN-order. These plots are at $R=6M$.}
\label{fig6}
\end{figure*}

\begin{figure*} 
\centering
\includegraphics[width=0.65\textwidth]{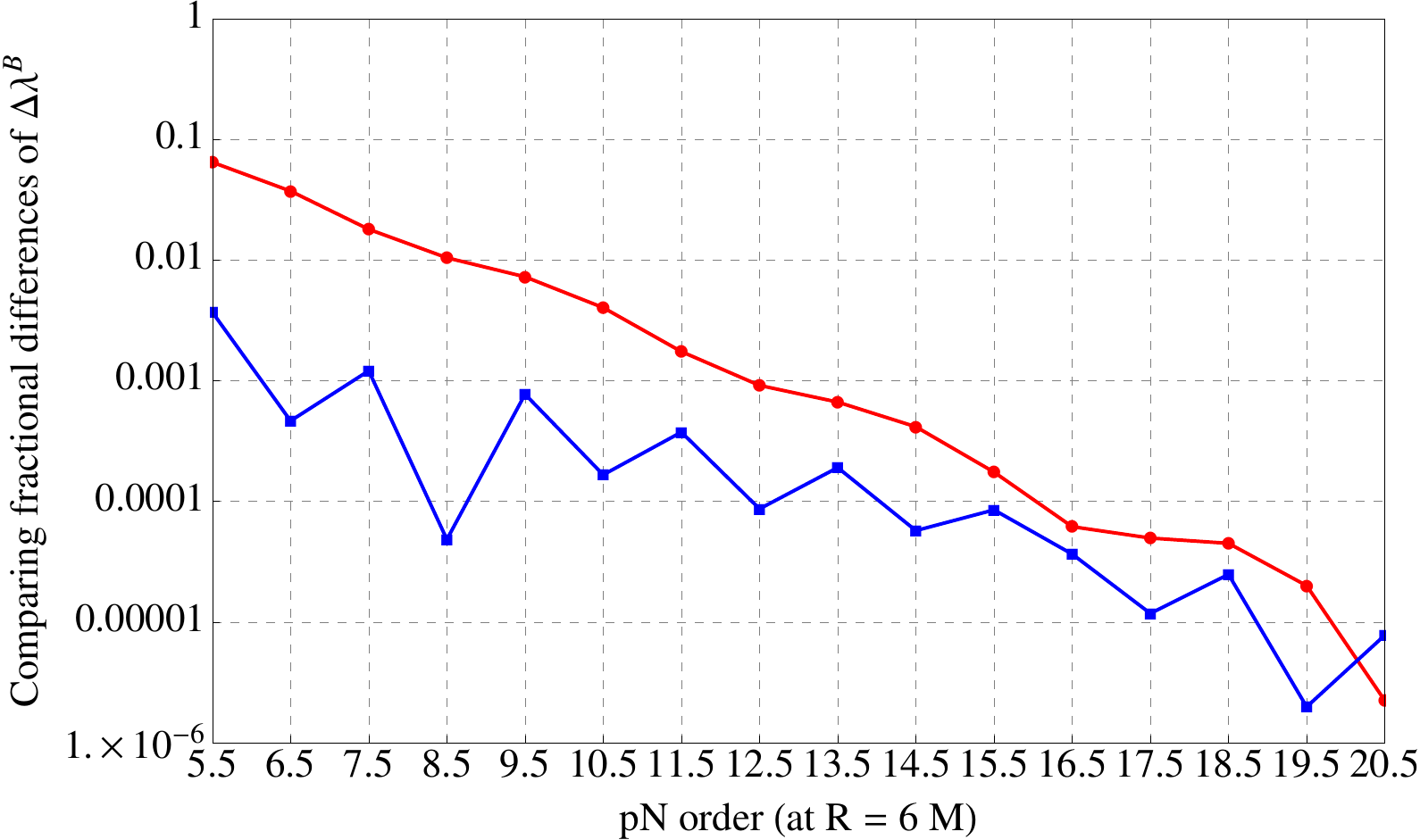}
\caption{Fractional difference between the  numerically calculated value and Taylor-summed pN series (represented by red line), and the fractional difference between the numerically calculated value and exponentially-resummed pN series (represented by blue line), for different pN-order. These plots are at $R=6M$.}
\label{fig6}
\end{figure*}

\begin{figure*} [htb]
\centering
\includegraphics[width=0.495\textwidth]{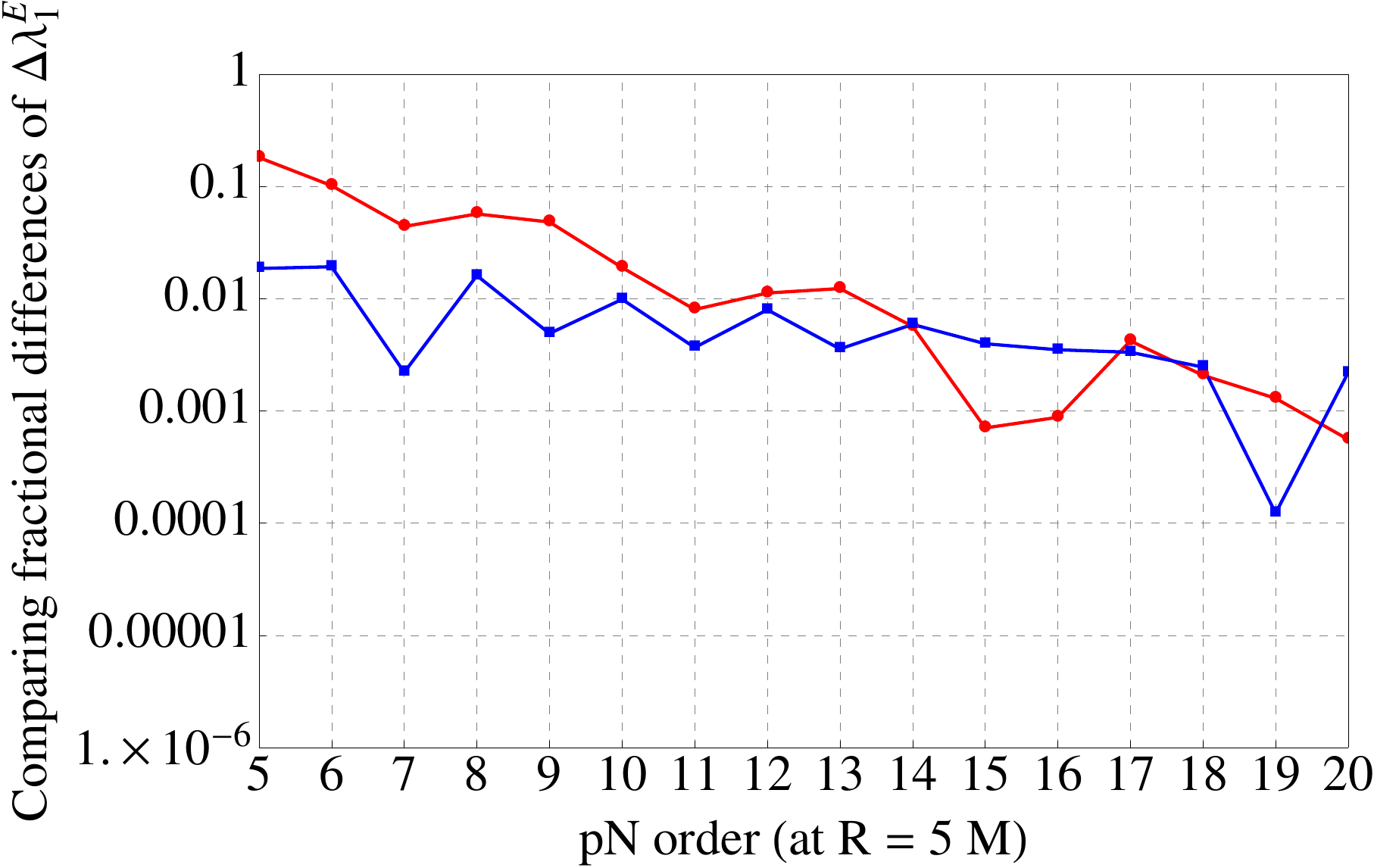}
\includegraphics[width=0.495\textwidth]{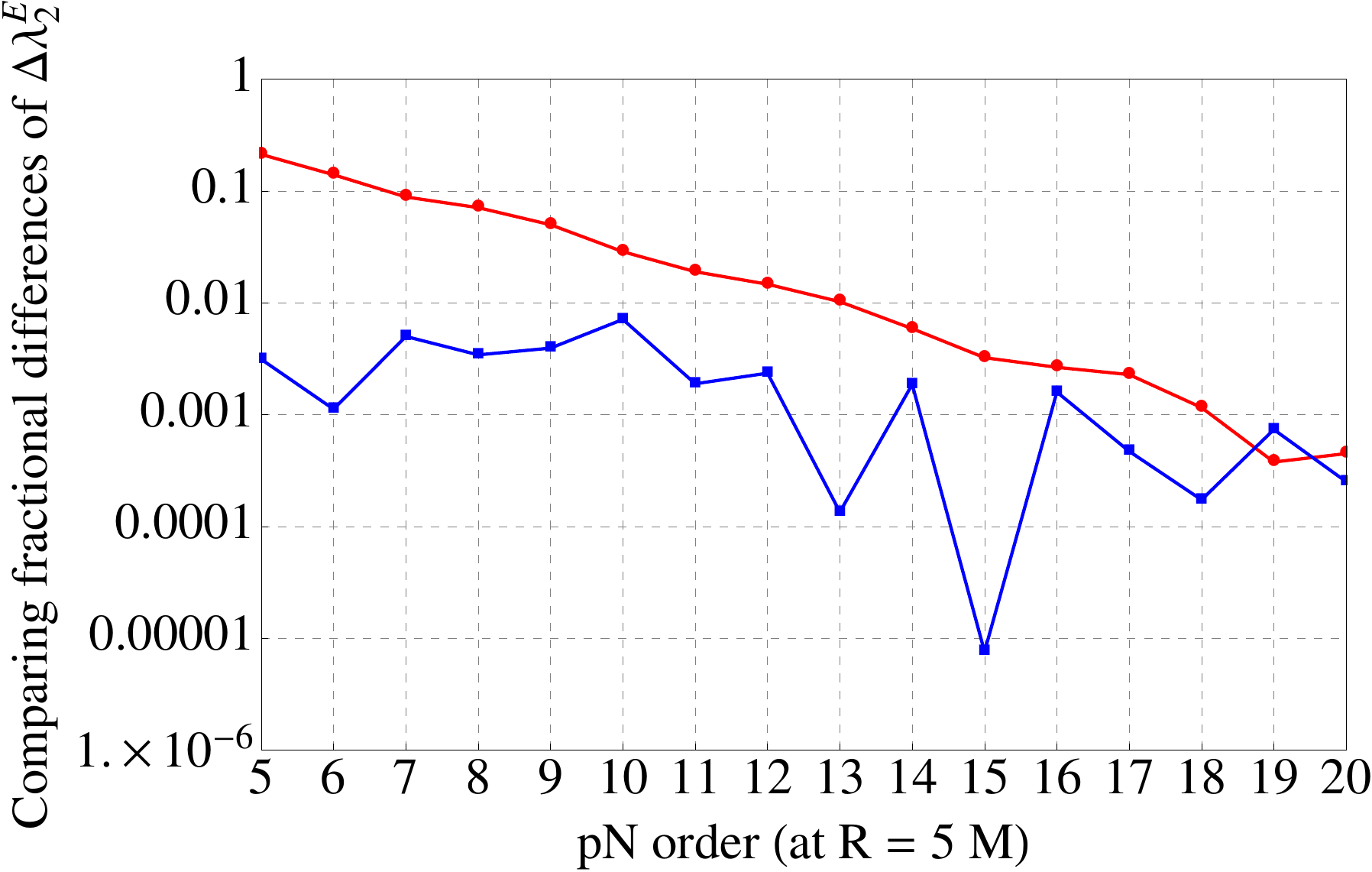}
\includegraphics[width=0.495\textwidth]{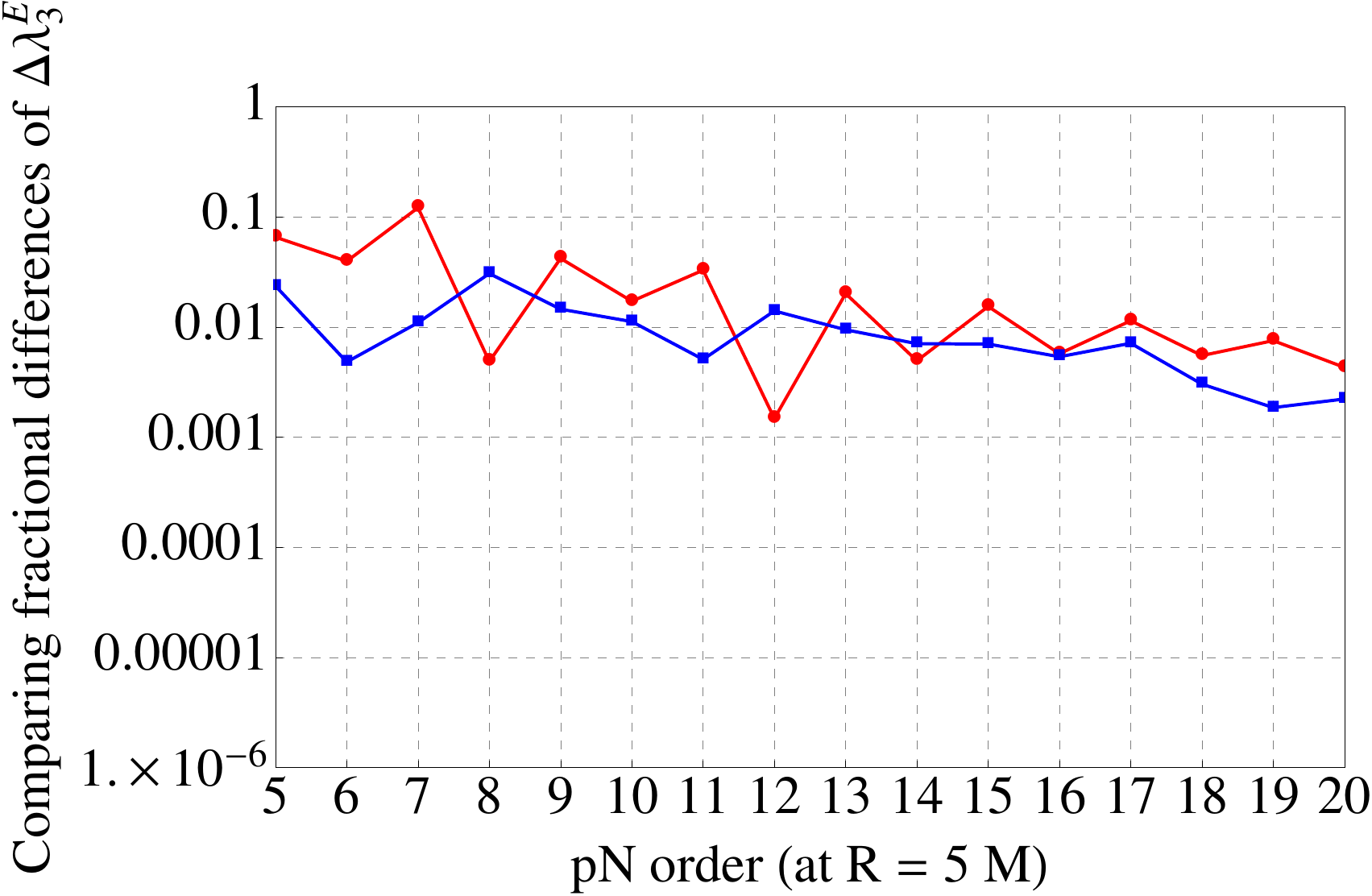}
\caption{Fractional difference between the  numerically calculated value and Taylor-summed pN series (represented by red line), and the fractional difference between the numerically calculated value and exponentially-resummed pN series (represented by blue line), for different pN-order. These plots are at $R=5M$.}
\label{fig7}
\end{figure*}

\begin{figure*}
\centering
\includegraphics[width=0.65\textwidth]{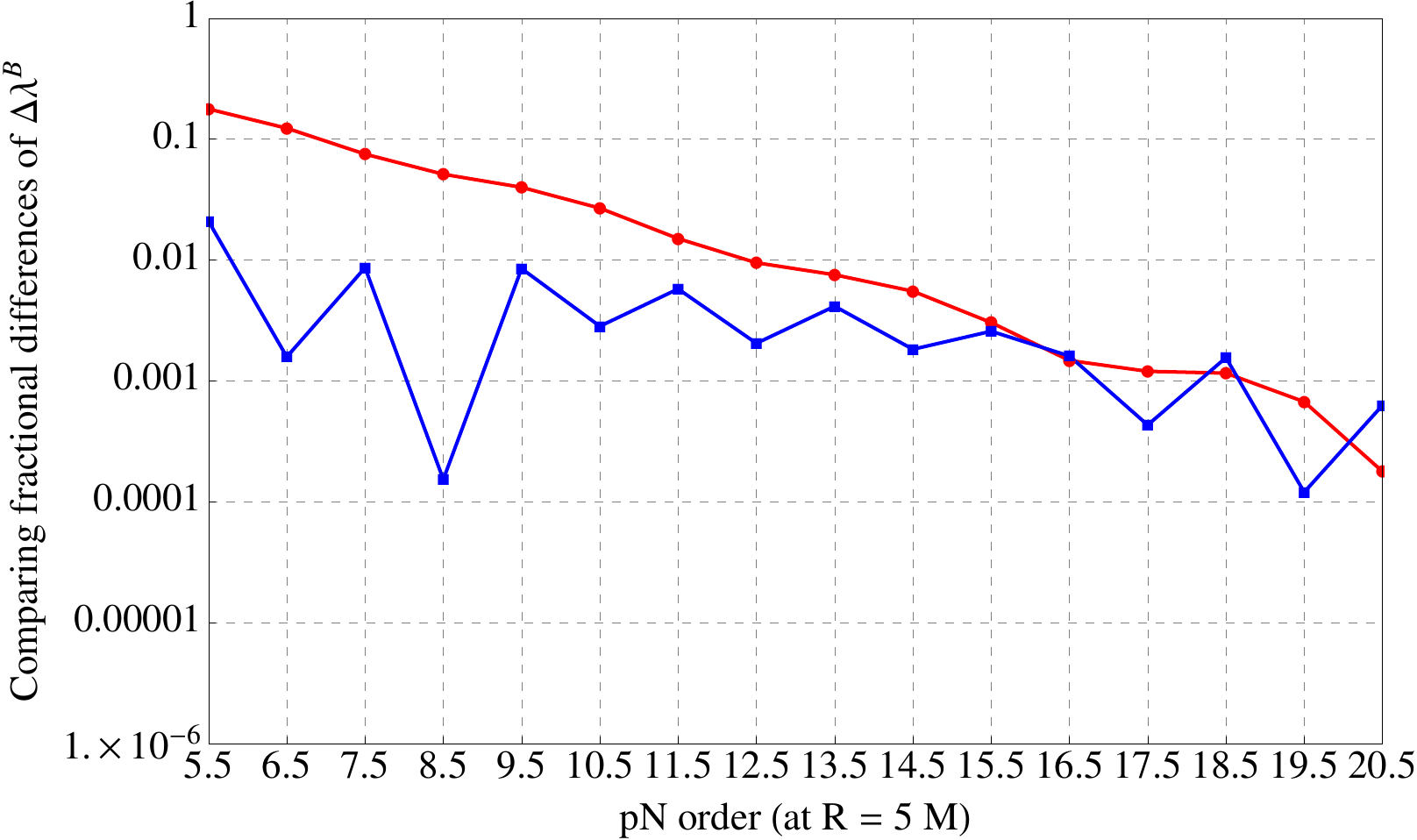}
\caption{Fractional difference between the  numerically calculated value and Taylor-summed pN series (represented by red line), and the fractional difference between the numerically calculated value and exponentially-resummed pN series (represented by blue line), for different pN-order. These plots are at $R=5M$.}
\label{fig7}
\end{figure*}

\begin{figure*} [htb]
\centering
\includegraphics[width=0.495\textwidth]{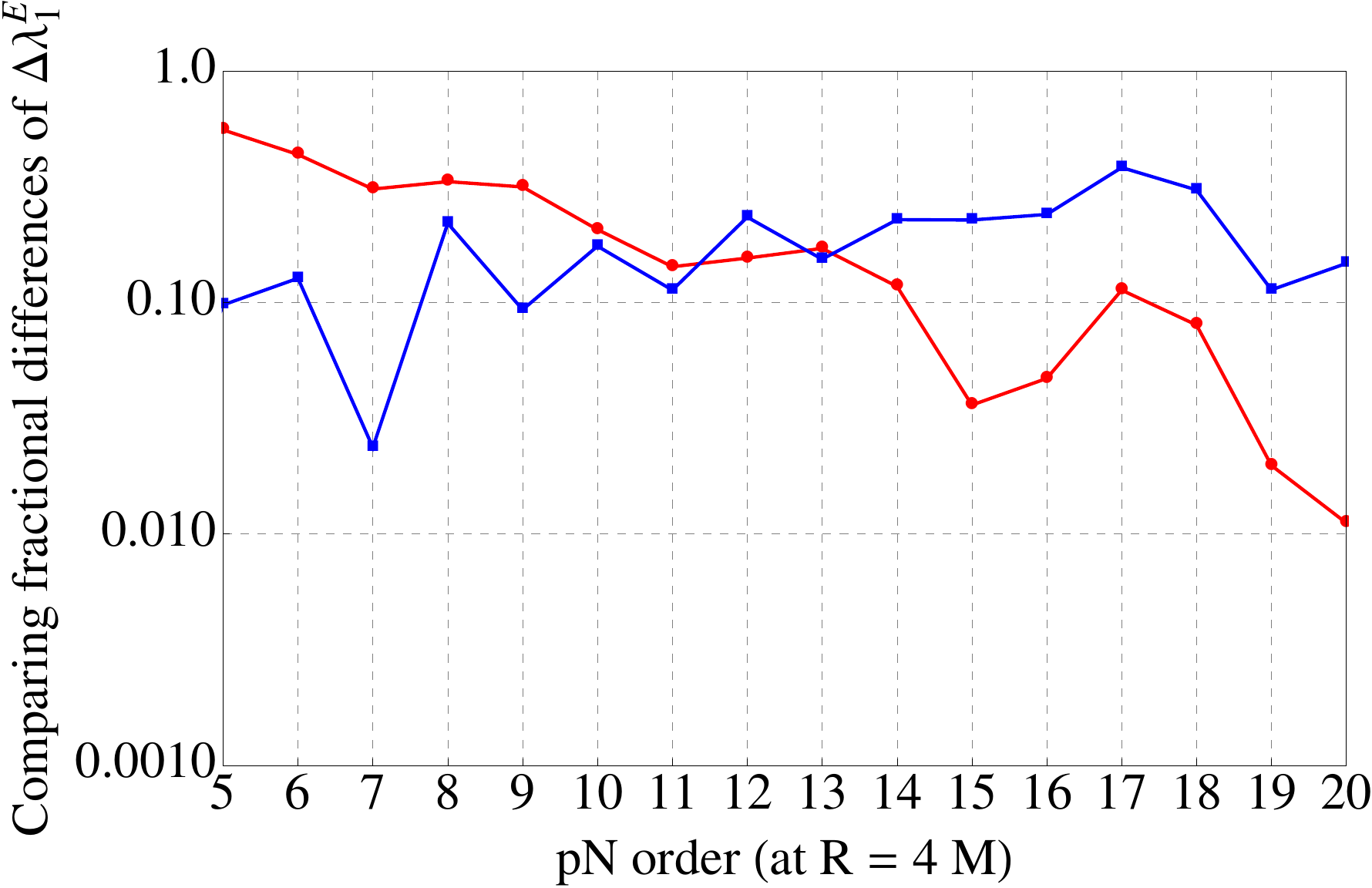}
\includegraphics[width=0.495\textwidth]{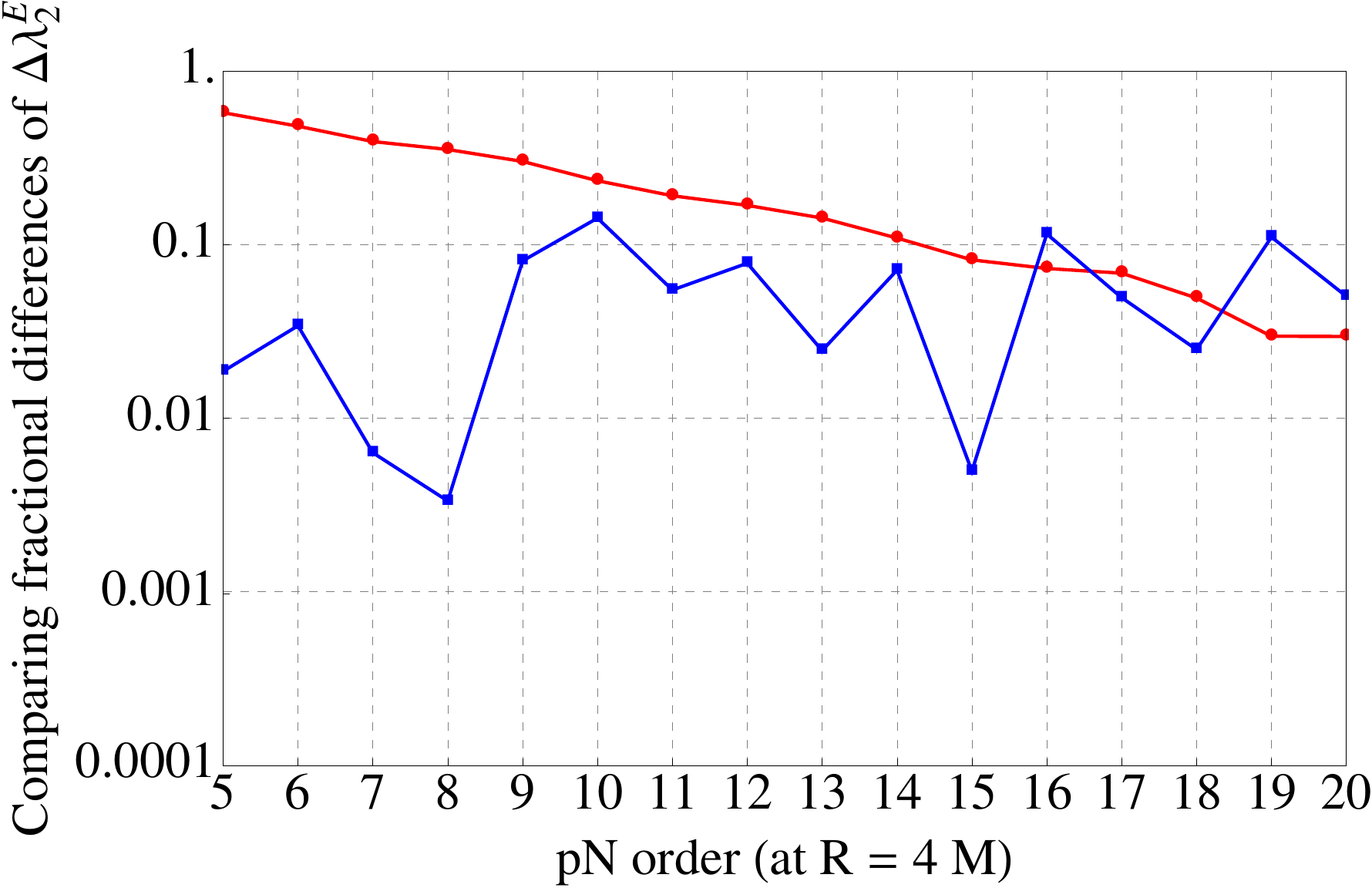}
\includegraphics[width=0.495\textwidth]{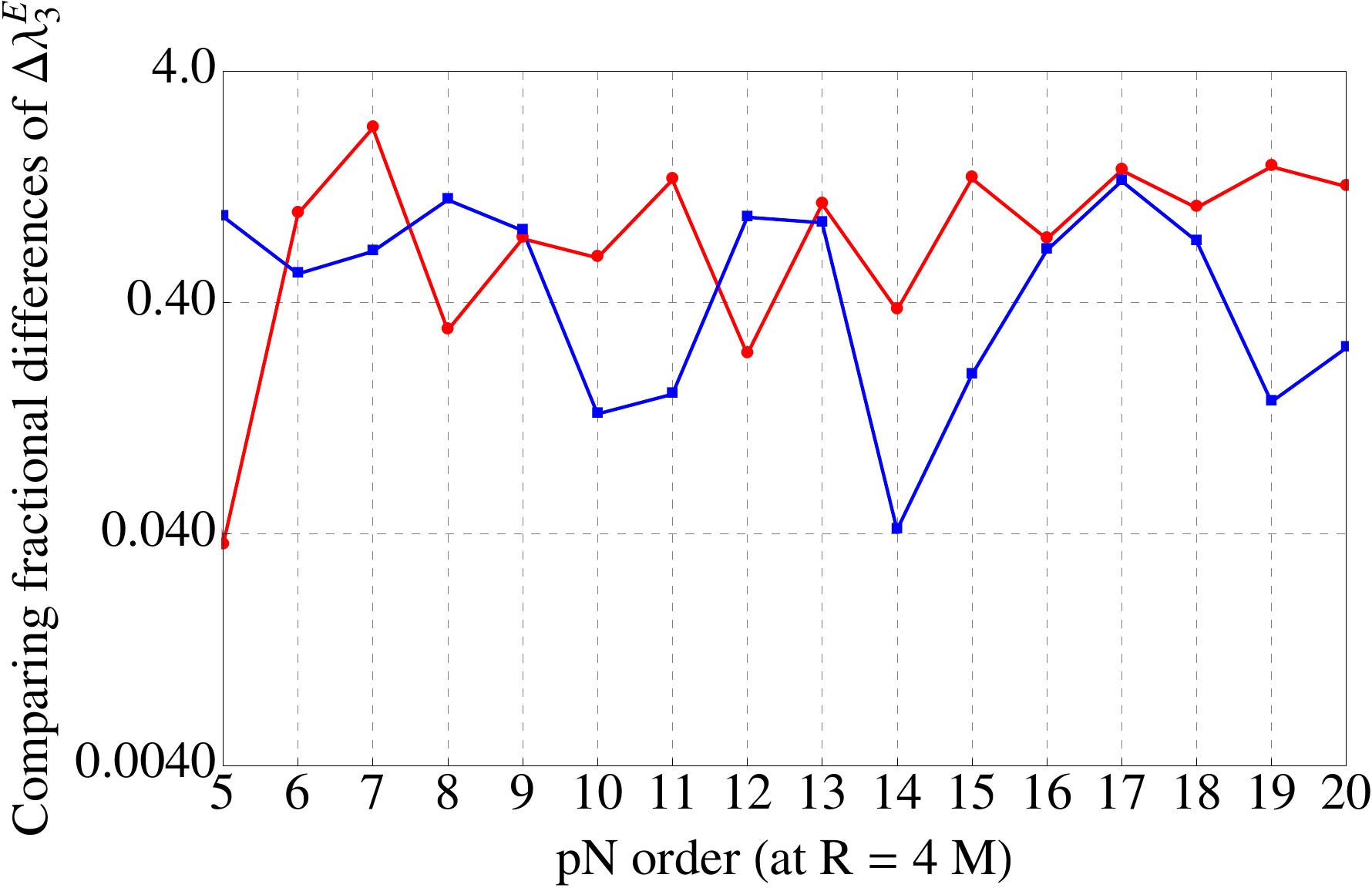}
\caption{Fractional difference between the  numerically calculated value and Taylor-summed pN series (represented by red line), and the fractional difference between the numerically calculated value and exponentially-resummed pN series (represented by blue line), for different pN-order. These plots are at $R=4M$.}
\label{fig8}
\end{figure*}

\begin{figure*} [htb]
\centering
\includegraphics[width=0.65\textwidth]{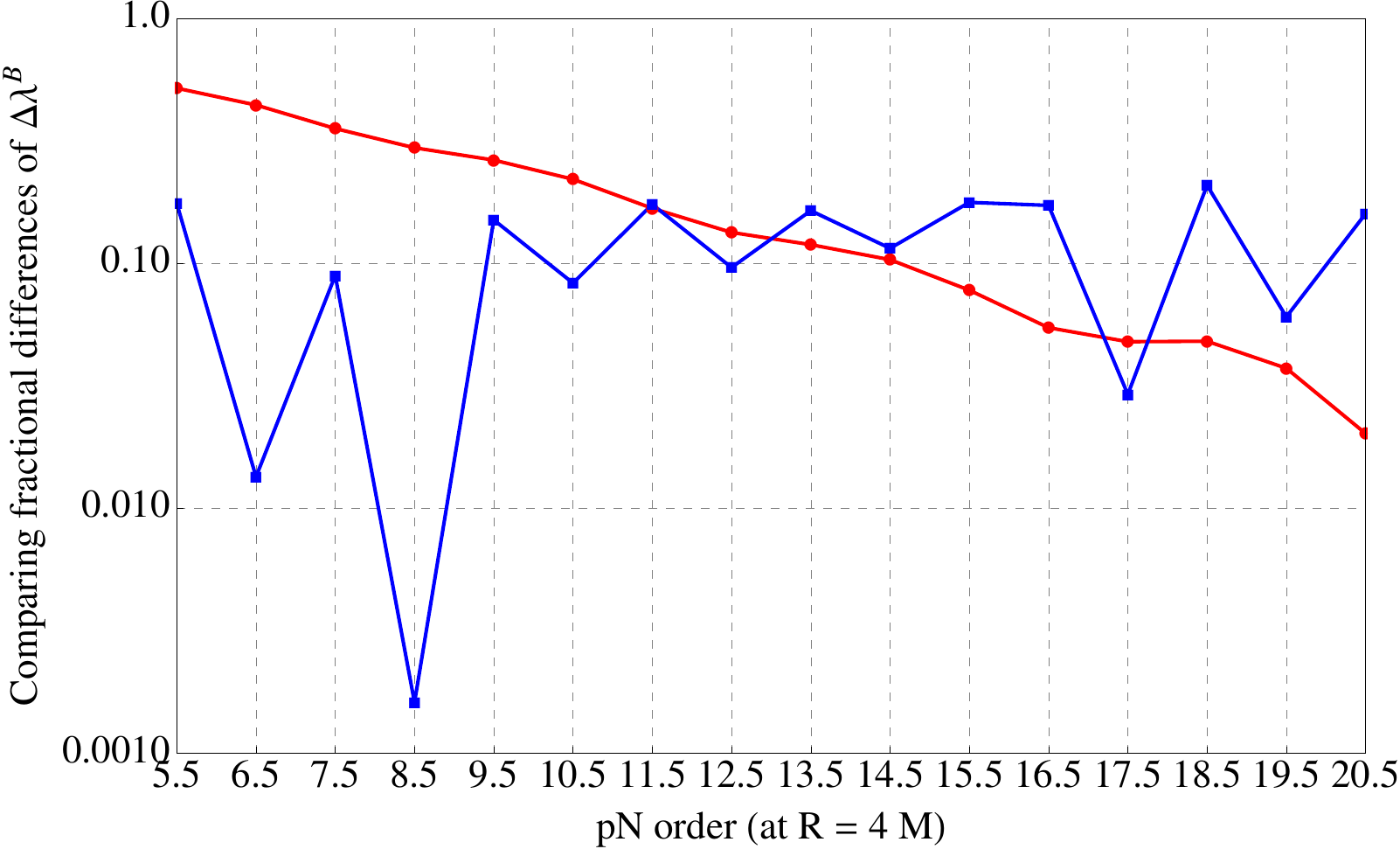}
\caption{Fractional difference between the  numerically calculated value and Taylor-summed pN series (represented by red line), and the fractional difference between the numerically calculated value and exponentially-resummed pN series (represented by blue line), for different pN-order. These plots are at $R=4M$.}
\label{fig8}
\end{figure*}


\end{document}